\documentclass[journal]{IEEEtran}
\usepackage{amsmath,amsfonts}
\usepackage{algorithmic}
\usepackage{algorithm}
\usepackage{array}
\usepackage[caption=false,font=normalsize,labelfont=sf,textfont=sf]{subfig}
\usepackage{textcomp}
\usepackage{stfloats}
\usepackage{url}
\usepackage{verbatim}
\usepackage{graphicx}
\usepackage{cite}
\usepackage{caption}
\usepackage{lipsum}
\usepackage{float} % Added for figure environment
\usepackage{xcolor}

\usepackage{tabu}    
\usepackage{booktabs}                  % only used for the table example
\usepackage{lipsum}                    % used to generate placeholder text
\usepackage{mwe}                       % used to generate placeholder figures

\usepackage{multirow}
\usepackage{makecell}
\usepackage{xcolor} % or \usepackage[table]{xcolor} if you need colored tables
\usepackage[normalem]{ulem}
\usepackage{mathptmx}                  % use matching math font

\begin{document}

\title{TransGI: Real-Time Dynamic Global Illumination With Object-Centric Neural Transfer Model}

\author{Yijie Deng$^*$, Lei Han$^*$, Lu Fang$^{\S}$
\thanks{Yijie Deng is with Tsinghua University and Tsinghua Shenzhen International Graduate School. E-mail: fightingdyj@gmail.com}% <-this % stops a space
\thanks{Lei Han is with Huawei Technology, Inc. E-mail: hanlei40@hisilicon.com.}
\thanks{Lu Fang is with Dept. of Electrical Engineering at Tsinghua University and also Beijing National Research Center for Information Science and Technology (BNRist), Beijing 100084. E-mail: fanglu@tsinghua.edu.cn}
\thanks{$*$: Equal Contribution. }
\thanks{$\S$: Correspondence Author}
\thanks{Manuscript received June 4, 2024; revised April 6, 2025.}}

\maketitle

\begin{figure*}[!t]
    \centering
    \includegraphics[width=1\linewidth]{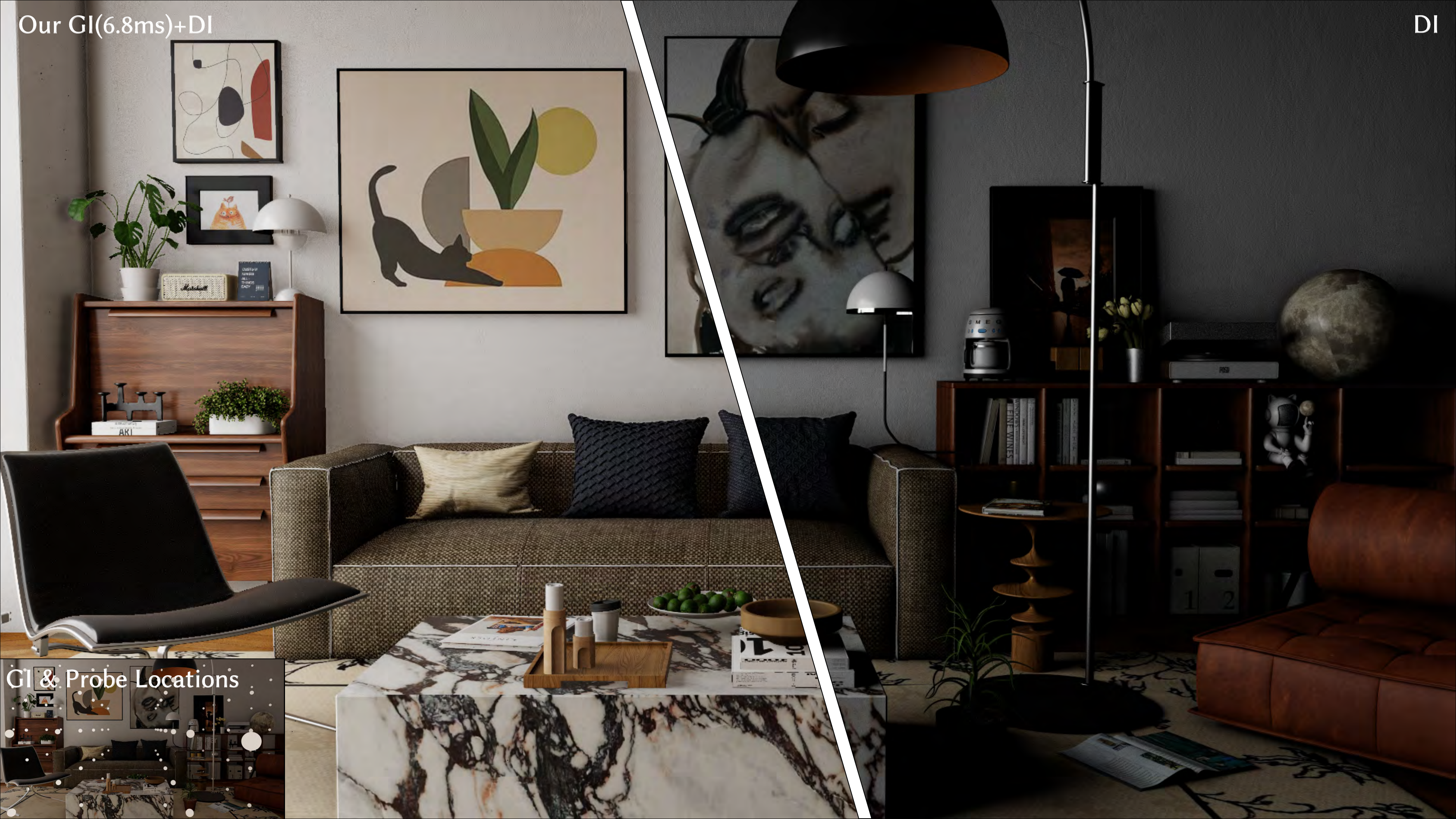}
    \caption{Left: Real-time global illumination result produced by our method, combined with path-traced direct illumination. Right: Path-traced direct illumination. Bottom left: The overall global illumination result and the probe locations of our method.}
    \label{fig:teaser}
\end{figure*}

\begin{abstract}
Neural rendering algorithms have revolutionized computer graphics, yet their impact on real-time rendering under arbitrary lighting conditions remains limited due to strict latency constraints in practical applications. The key challenge lies in formulating a compact yet expressive material representation.
To address this, we propose \textit{TransGI}, a novel neural rendering method for real-time, high-fidelity global illumination. It comprises an object-centric neural transfer model for material representation and a radiance-sharing lighting system for efficient illumination.
Traditional BSDF representations and spatial neural material representations lack expressiveness, requiring thousands of ray evaluations to converge to noise-free colors. Conversely, real-time methods trade quality for efficiency by supporting only diffuse materials.
In contrast, our object-centric neural transfer model achieves compactness and expressiveness through an MLP-based decoder and vertex-attached latent features, supporting glossy effects with low memory overhead.
For dynamic, varying lighting conditions, we introduce local light probes capturing scene radiance, coupled with an across-probe radiance-sharing strategy for efficient probe generation.
We implemented our method in a real-time rendering engine, combining compute shaders and CUDA-based neural networks. Experimental results demonstrate that our method achieves real-time performance of less than 10 ms to render a frame and significantly improved rendering quality compared to baseline methods.
\end{abstract}


\section{Introduction}
\IEEEPARstart{R}{eal}-time global illumination (GI) is crucial for achieving high visual quality and an immersive interactive experience in video games and other industrial designs. However, contemporary methods for high-fidelity global illumination incur substantial computational costs, necessitating trade-offs between quality and temporal efficiency. To clarify terminology, in this work we follow the convention used in DDGI~\cite{majercik2019dynamic}, where global illumination specifically refers to indirect lighting.

A widely adopted approach for real-time global illumination is the probe-based lighting system, exemplified by dynamic diffuse global illumination (DDGI) \cite{majercik2019dynamic}. This technique achieves real-time performance by minimizing computations from the per-pixel level to sparsely scattered light probes, simultaneously simplifying materials to diffuse-only representations. While this approach offers computational efficiency, the oversimplification of light probes and object materials imposes severe limitations, hindering support for complex materials and realistic lighting scenarios.

Recently, neural rendering algorithms have demonstrated the potential to revolutionize computer graphics by learning representations and complex light-material interactions from data. Real-time neural appearance models \cite{zeltner2023real} have found a way to represent complicated materials in a unified manner using small multi-layer perceptrons, but they still operate within the traditional path-tracing pipeline, which cannot eliminate Monte-Carlo noise. Neural precomputed radiance transfer (Neural-PRT) \cite{RBRD22} leverages the instant lighting concept from PRT \cite{sloan2023precomputed} and encodes the scene radiance transfer in a neural network. By decoding the transfer vectors of the viewing frame and the light vectors from the environment maps, novel views can be rendered by combining these two vectors. However, the models employed in Neural-PRT are fitted per-scene and fixed at the scene scale. Consequently, they can only be utilized for static scene geometries, lacking the ability to support object-level dynamics such as movement and rotation. Real-time neural radiance cache (NRC) \cite{10.1145/3450626.3459812} proposed to use an online-trained neural network to memorize the global illumination, but it still requires high-quality path-traced results for training, which is not feasible on low-end devices.

To address these limitations and provide a more robust and flexible global illumination algorithm, we propose \textit{TransGI} to achieve high-quality global illumination in real-time with the proposed object-centric neural transfer model and the real-time radiance-sharing lighting system. Our rendering process is built upon Precomputed Radiance Transfer (PRT) \cite{sloan2023precomputed}, where an object's shaded response to light sources is precomputed and compressed into a transfer function. This transfer function is represented in a linear basis, typically using spherical harmonics for a single evaluated outgoing direction, usually called \textit{transfer coefficients}. It maps the incoming radiance to the outgoing radiance. The incoming light is dynamically sampled and represented using the same basis, usually called \textit{light coefficients}. Consequently, the rendering process is reduced from evaluating the light integral to a simple dot product between transfer coefficients and light coefficients. This approach offers significant computational savings during real-time rendering.

% \textcolor{red}{Specifically, we model the rendering process as a precomputed transfer function following~\cite{sloan2023precomputed}}, where the object's shaded response to lights is calculated as a transfer function, mapping the imcoming to outgoing radiance. The transfer function is a view dependent spatially variant 4D function that modeled as neural networks.

% \textcolor{red}{Following the tradition in PRT\cite{sloan2023precomputed}, where an object's shaded response to lights is calculated and compressed as a \textit{transfer function}, mapping the incoming to outgoing radiance. Our method also models the object's shaded response as transfer functions, which is used to decode transfer vectors  decomposes the rendering process into estimating light coefficients and object transfer coefficients. The light coefficients represent the light transfer functions of the scene, while the transfer coefficients encode the material properties of the objects.}

% \textcolor{red}{{PRT-based approaches are widely used in offline baking-based rendering systems thanks to its efficiency and visual quality, yet are hardly employed in real-time rendering applications due to the following limitations: }}

% \begin{itemize}
%     \item Representation Capability. Complex glossy materials takes GBs storage overhead.
%     \item Generalization to dynamic scenes.
%     \item Far light assumption limits its application in scenes with complex lighting conditions. 
% \end{itemize}

However, traditional PRT approaches exhibit inherent limitations that restrict their applicability in contemporary real-time rendering applications: (1) Static scenes. PRT operates by precomputing and storing the light transport information for a given scene geometry. Consequently, the scene geometry must remain static during rendering, as any changes would invalidate the precomputed data. (2) Limited representational capabilities. To achieve real-time performance, PRT simplifies the representation of materials and lighting models. This simplification constrains the types of materials and lighting scenarios that can be accurately depicted. (3) Restricted lighting dynamics. PRT typically assumes a global distant light source, which may not be suitable for scenes with complex, dynamic lighting conditions.

To address these issues, our method solves these problems by: 

\begin{itemize}
    \item Representing radiance transfer at object-scale instead of scene-scale, thereby supporting object-level dynamics such as object movement, addition, and removal.
    \item Utilizing neural networks and latent features to compress densely-sampled object transfer functions, enabling the representation of more realistic and complex materials.
    \item Employing local light probes to store scene lighting information, facilitating dynamic lighting scenarios, such as moving light sources and rapidly changing lighting conditions.
\end{itemize}

% In this paper, we attack these problems with the following improvements: A. Neural representation with little memory overhead. B. Object-centric representation for dynamic scenes. C. Distributed light probes for novel view synthesis.

% Traditional PRT is a simplified lighting system which represents transfer coefficients using only 1D vector for diffuse materials and 2D matrix for plain glossy materials with no textures. And it only supports distant lights. But it can achieve real-time and noise-free rendering because the rendering process reduces from endless per-ray integration to simple multiplication of two well-established frequency coefficients. We hope to leverage its advantage of real-time rendering but extend its capability to support more complex materials and lighting scenarios.

Specifically, our object-centric neural transfer model is constructed by first densely sampling the object transfer function and then training and compressing it within a neural network and vertex-attached latent features. For a high-fidelity representation of the object transfer, it should be fully sampled at the object surface positions and outgoing directions. However, this poses a severe challenge for storage if we store the raw sampled transfer coefficients for querying and interpolation. For instance, a sampling density of 100 points per area and 1000 directions per point for a box of one square area would require over 20GB of storage for a 300-byte basis, which is not feasible in real-time rendering pipelines. To address this issue, we propose compressing the densely-sampled transfer function using a neural network and vertex-attached latent features. The neural network takes the vertex latent vectors, the outgoing direction, and necessary G-Buffers (Geometry Buffers) as input and outputs the transfer coefficients. After training, the neural network can achieve a very high compression rate, requiring less than 1MB of storage, while the per-vertex latent features are only a 1D vector, also occupying a small space in storage.

% To support more complex materials, the transfer coefficients should be stored with high fidelity, which poses a challenging problem of storage overhead. As the transfer coefficients vary at different locations and outgoing directions, the storage overhead is proportional to the sampling density of positions and outgoing directions. For example, a sampling density of 100 points per-area and 1000 directions per-point for a box of one area would require over 20GB of storage for a 300-byte basis, which is not feasible in real-time rendering pipelines. To address this issue, we propose to compress the transfer coefficients using a neural network and vertex-attached latent features. The neural network takes the vertex latent vectors, the outgoing direction, and necessary G-Buffers as input and outputs the transfer coefficients. The neural network is trained with densely-collected transfer coefficients and can achieve a really high compression rate, less than 1MB, while the per-vertex latent features are only a 1D vector, ensuring high-fidelity rendering.

As for the proposed real-time radiance-sharing lighting system, it is composed of sparsely-scattered light probes, generated in real-time with our across-probe radiance-sharing strategy. Similar to DDGI \cite{majercik2019dynamic}, we place sparsely-scattered light probes in the scene, and each probe is equipped with a path-traced irradiance map to capture the local radiance energy. However, as the color of each pixel in the irradiance map is computed independently via path-tracing, the computational burden increases linearly with the number of light probes and the resolution of the irradiance map, hindering real-time efficiency. To address this issue, we propose the across-probe irradiance sharing strategy, where a probe can share the per-ray radiance energy of its neighboring probes, and the high-resolution irradiance map can be computed by rasterizing the shared shaded points. This approach allows for the efficient computation of high-resolution irradiance maps by leveraging the radiance information from neighboring probes, reducing the computational burden associated with independent path-tracing for each probe. The irradiance map is then projected onto a suitable basis (e.g., spherical harmonics) to obtain light coefficients for further rendering. 

By combining sparsely-scattered light probes with the across-probe irradiance sharing strategy, our real-time radiance-sharing lighting system enables efficient computation of high-quality, dynamic lighting while maintaining real-time performance.

The proposed algorithm is implemented in a real-time rendering engine Falcor\cite{kallweit22falcor} to verify its efficacy. This involves seamlessly integrating compute shaders and CUDA-based neural networks\cite{tiny-cuda-nn}. The whole system is optimized to run on portable devices and performs significantly better than DDGI, which cannot represent the view-dependent GI of glossy surfaces. We have also conducted extensive experiments to validate the real-time performance and high-fidelity rendering of our method, achieving high efficiency of less than 10 ms to render a frame and significantly higher quality than the baseline method DDGI.

Figure \ref{fig:teaser} showcases the global illumination rendering results produced by our method. As demonstrated, our approach generates real-time, visually smooth, and physically plausible global illumination effects, significantly enhancing the realism of lighting in indoor environments.

In summary, our contributions are:

\begin{itemize}
    % \item A compact and high-fidelity neural frequency representation for complicated materials[EXPLAIN HOW COMPLICATED!].
    \item A compact \textit{object-centric neural transfer model} enabling high-fidelity dynamic global illumination of complex geometry and materials with minimal memory overhead. 
    % \item \textcolor{blue}{An object-centric neural appearance representation that models surface materials at the object-level. Unlike the existing neural precomputed radiance transfer (PRT) method \cite{RBRD22} that operates at the scene-scale, our object-level representation enables more generalizable operations such as near-field relighting, object movement, object addition, and removal.}
    \item A real-time probe-based lighting system leveraging the \textit{across-probe radiance-sharing} strategy for efficient computation of high-quality, dynamic lighting.
    \item A real-time implementation for seamlessly integrating neural material models within a widely adopted rendering engine Falcor\cite{kallweit22falcor}, demonstrating the feasibility and performance of the proposed techniques in real-world applications.
\end{itemize}
% Head 1
\section{Related Work}
Global illumination is crucial for achieving realistic lighting in computer graphics, yet it remains a challenging problem due to the following reasons: (1) It is difficult to support multi-layered complex materials at interactive frame rates. (2) It is challenging to support dynamic lighting conditions in real-time.

Physically-based rendering methods, featured by path tracing \cite{veach1997metropolis,kelemen2002simple,bitterli2019selectively,veach1995bidirectional}, are capable of achieving realistic rendering quality through Monte Carlo integration of the rendering equation \cite{kajiya1986rendering}. However, these methods are computationally expensive and produce noisy rendering results at low sample counts.

To achieve real-time global illumination, various approaches have been proposed, primarily based on the following two simplifications:
\begin{itemize}
    \item Material simplifications.
    \item Dynamics Restrictions.
\end{itemize}
We review the existing real-time global illumination methods and compare our proposed method with them in Section \ref{subsec:re1}, highlighting how our approach overcomes the limitations of material simplifications and dynamics restrictions. As neural networks have been increasingly adopted in computer graphics for representing complex materials, we review the neural material representation methods in Section \ref{subsec:re2} and compare them with our neural transfer model, which aims to achieve real-time rendering performance without compromising on representing object materials and dynamic lighting conditions. 

\subsection{Real-time Global Illumination}
\label{subsec:re1}
% \textbf{Precomputation}.

To achieve real-time performance, most existing global illumination methods impose restrictions on light transport, primarily supporting only diffuse materials. The most widely adopted technique in industry is Dynamic Diffuse Global Illumination (DDGI) \cite{majercik2019dynamic}, which represents the irradiance field using scattered light probes, consequently limiting light propagation to diffuse paths. Similar approaches, such as \cite{greger1998irradiance,mcguire2017real,silvennoinen2017real,ritschel2009approximating,nichols2009hierarchical}, also constrain their computation to diffuse materials or limited light bounces, sacrificing support for more complex light interactions. While Neural Radiance Cache (NRC)\cite{10.1145/3450626.3459812} employs a neural network to efficiently store and retrieve the rendered scene radiance, its performance is suboptimal when handling glossy materials due to the complexity of accurately modeling specular reflections and their intricate light transport interactions. To overcome the limitations of existing methods in supporting complex materials, our approach employs a compact neural representation that encodes the intricate transfer properties of opaque materials. This neural transfer model captures glossy effects while maintaining real-time performance, thereby enabling more realistic rendering of sophisticated materials.

To support more complicated materials, some scene dynamics and lighting dynamics are sacrificed in some existing global illumination methods. Precomputed Radiance Transfer (PRT)\cite{sloan2023precomputed} precomputes the radiance transfer function of the scene to handle glossy materials and achieve efficient rendering. However, it is limited to distant lighting and does not support scene dynamics such as object movement or addition/removal, as the transfer function treats all objects as a whole. A similar approach boosted with neural networks, Neural-PRT\cite{RBRD22}, fits the scene radiance transfer in a neural network and decodes the transfer vectors and lighting vectors with a fitted decoder. However, it is trained per scene, lacking generalizability. Other precomputation methods\cite{ng2003all,liu2004all,10.1145/1179352.1141980,10.1145/1138450.1138456,10.1145/1618452.1618486} use different representations for complex materials but also do not support scene dynamics. \cite{rodriguez2020glossy} reproduces accurate glossy effects with sparse glossy light probes but requires hours of baking time and supports only static scenes. \cite{diolatzis2022active} supports some scene dynamics by training a neural network to store the radiance variance according to changing dynamic parameters, but it also requires lengthy training time and can only handle limited scene dynamics. To support flexible rendering while being efficient in all procedures, we make our neural representations object-centric, enabling scene dynamics such as object movement, removal, and addition. In contrast to PRT-based methods that only adapt to distant lighting, we use local light probes as in DDGI\cite{majercik2019dynamic} to support local and variant light changes. The probes can also be generated in real-time with our across-probe radiance-sharing strategy.

Lumen\footnote{https://advances.realtimerendering.com/s2022/SIGGRAPH2022-Advances-Lumen-Wright\%20et\%20al.pdf}, a prominent rendering technique extensively employed in video game development, exhibits high performance but necessitates high-end hardware for optimal frame rates and increased storage capacity to accommodate essential rendering caches. In contrast, our approach leverages lightweight neural representations, comprising solely of vertex-attached 1D vectors and a light MLP as the transfer decoder, thereby mitigating device requirements.

Overall, our method achieves both complex material representation and dynamic rendering through an object-centric neural transfer model and an across-probe radiance-sharing strategy for probe generation.

\subsection{Neural Material Representation}
\label{subsec:re2}
Neural networks have been increasingly adopted in computer graphics for representing complex materials, aiming to improve rendering quality and achieve more compact representations.

Several works have utilized neural networks for material representation. \cite{yan2017bssrdf} employed a neural network to convert fur parameters into a participating medium, simplifying the simulation. \cite{thies2019deferred} introduced neural textures and used them within deferred rendering. \cite{rainer2019neural} proposed an autoencoder framework to compress BTF slices, while \cite{fan2023neural} represented BTFs using neural biplanes and a general decoder for radiance decoding from the biplanes. \cite{rematas2016deep,maximov2019deep} applied CNNs to regress 2D or 4D maps that express the appearance of real objects. \cite{kuznetsov2019learning} used Generative Adversarial Networks (GANs) to generate reflectance functions perceptually similar to synthetic or measured input data, rendering them through partial evaluation of the generator network. \cite{10.1145/3450626.3459795} proposed a neural method for representing and rendering various material appearances at different scales with neural textures. \cite{10.1145/3528233.3530732,guo2023metalayer,zeltner2023real} focused on representing complex multi-layer materials using neural networks. Notably, \cite{zeltner2023real} encodes the BRDF model in small MLPs and supports multi-layer normal maps for complicated materials.
However, most of these methods are limited to offline rendering, except for \cite{zeltner2023real}, which demonstrated real-time implementation in a real-time rendering engine. Nevertheless, \cite{zeltner2023real} still operates within the traditional path-tracing paradigm, suffering from Monte Carlo noise and requiring post-processing techniques for noise-free rendering results.

Our objective is to capitalize on the simplicity and computational efficiency of the lightweight neural representation employed in \cite{zeltner2023real} for real-time rendering. However, instead of modeling the spatial Bidirectional Reflectance Distribution Function (BRDF), we aim to model the transfer properties of materials across the entire frequency spectrum, facilitating more informative encoding and noise-free rendering. To maintain high computational efficiency, we strive to keep the object representation as simple as possible, opting for 1D vertex latent features, while leveraging a lightning-fast Multilayer Perceptron (MLP) neural decoder implemented in CUDA for accelerated performance.

\section{Overview}
\label{sec:overview}

\begin{figure*}
    \centering
    \includegraphics[width=1\linewidth]{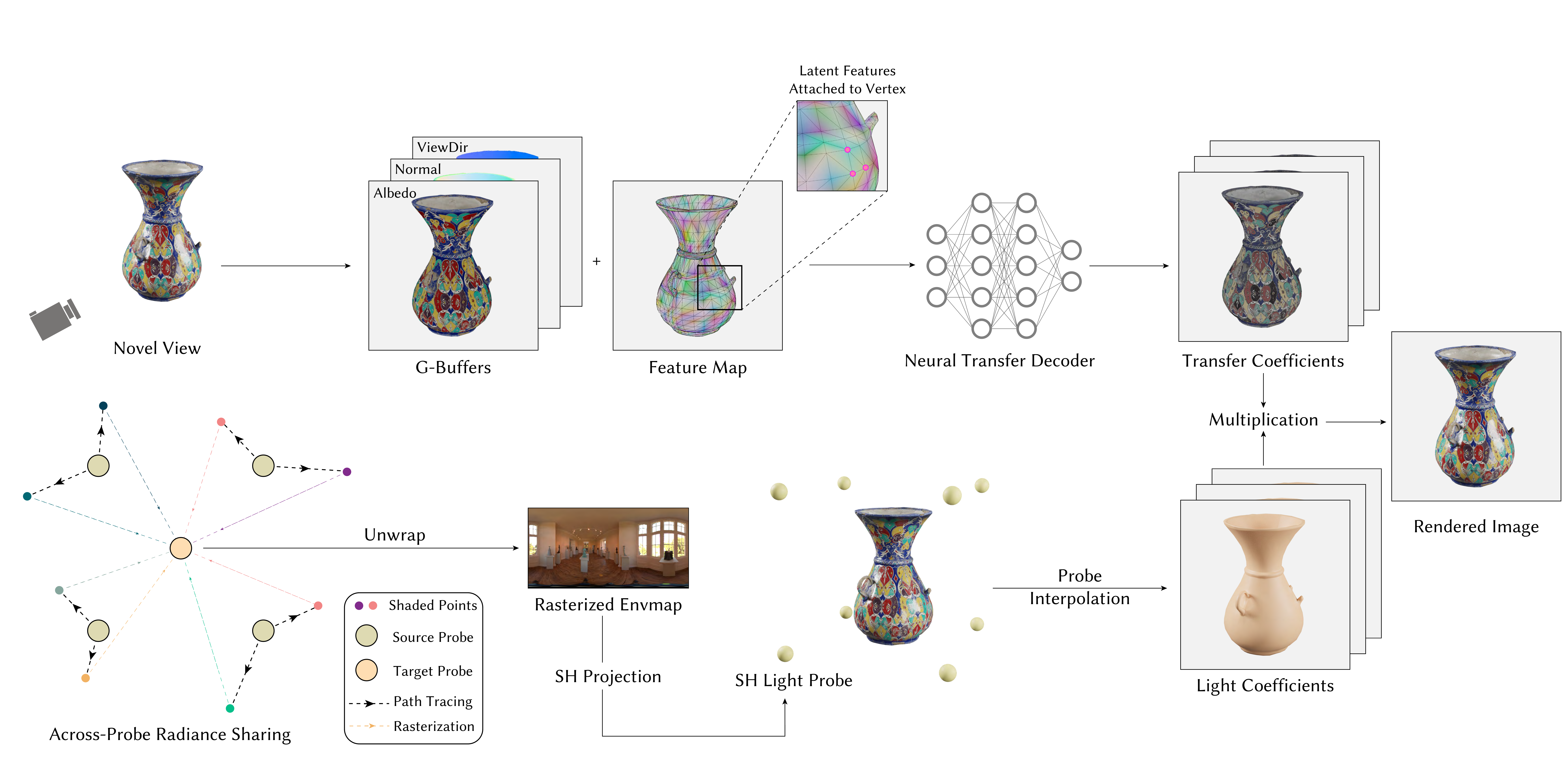}
    \caption{The pipeline of our method. Top row: A novel view is specified, and the corresponding G-Buffers (albedo, surface normal, and view direction) and the feature map are generated within the rendering engine. The feature map is interpolated with per-vertex latent features using barycentric coordinates. These generated maps are then fed into the neural transfer decoder to output the transfer coefficients. Bottom row: The irradiance map of a target probe is rasterized using its and its neighboring source probes' shaded points, which is then projected into the frequency domain using spherical harmonics basis, referred to as an SH Light Probe. To shade a surface point, the light coefficients are interpolated using conditional trilinear interpolation. Finally, the rendered image is produced by multiplying the transfer coefficients and light coefficients.}
    \label{fig:pipeline}
\end{figure*}

% Figure \ref{fig:pipeline} illustrates the entire pipeline, where the rendering process is decomposed into two workflows: decoding transfer coefficients and generating light coefficients. These two rendering components are analogous to the notations used in the traditional PRT \cite{sloan2023precomputed}. The meanings of these components are as follows:
% \begin{itemize}
% \item Transfer coefficients: The transfer coefficients represent the material's transfer properties, encoding how the material interacts with light at different frequencies. They capture the complex behavior of the material's reflectance properties, describing how the material absorbs, scatters, and reflects light.
% \item Light coefficients: The light coefficients represent the incoming radiance energy from all incident angles. They are multiplied by the transfer coefficients to compute the final shading color, effectively combining the material's response to light with the incident illumination to produce the desired shading effect.
% \end{itemize}

The whole pipeline is illustrated in Figure \ref{fig:pipeline}, where the rendering process is decomposed into two workflows: decoding transfer coefficients and generating light coefficients. These two rendering components are analogous to the notations used in the traditional PRT\cite{sloan2023precomputed}, and we explain their meanings below:
\begin{itemize}
    \item Transfer coefficients: These coefficients represent the material's transfer properties, encoding how the material interacts with light at different frequencies, capturing the complex behavior of the material's reflectance properties.
    \item Light coefficients: These coefficients represent the incoming radiance energy from all incident angles, which are multiplied by transfer coefficients to get the final shading color.
\end{itemize}

To better understand the mathematical formulation of the proposed method, we provide a brief introduction to Precomputed Radiance Transfer (PRT) in this section. Subsequently, we illustrate the process of decoding transfer coefficients in Section \ref{sec:appearance} and generating light coefficients in Section \ref{sec:probe}, respectively.

Our objective is to facilitate real-time global illumination that supports high-fidelity rendering, particularly for intricate reflective materials. This entails accurately reproducing the outgoing radiance $L_o(x_o,\omega_o)$ leveraging both the incoming radiance $L_i(x_o)$ and the surface material response $f(x_o, \omega_o)$:
\begin{equation}
\label{eq:01}
    L_o(x_o,\omega_o) = F(L_i(x_o), f(x_o, \omega_o)),
\end{equation}
where $x_o$ represents the shaded point's position, $\omega_o$ is the outgoing direction, and $F$ denotes the function that combines the incoming radiance and surface material response to yield the shaded color.

Traditionally, Equation (\ref{eq:01}) is evaluated by representing the surface material response as the spatially varying bidirectional reflective distribution function (SVBRDF) $f(x,\omega_i,\omega_o)$ and the incoming radiance as the spatially varying per-ray radiance $L_i(x_o, \omega_i)$:
\begin{equation}
\label{eq:02}
    L_o(x_o,\omega_o) = \int_\Omega f(x,\omega_i,\omega_o)L_i(x_o, \omega_i)(\omega_i \cdot n) d\omega_i,
\end{equation}
where $\omega_i$ represents the incident light direction, $n$ is the surface shading normal direction, and $\Omega$ is the upper hemisphere defined by $n$. 

However, rendering reflective materials using SVBRDF and its variants often results in noisy outcomes due to two primary reasons. Firstly, inadequate samples per pixel can lead to significant Monte Carlo sampling noise. Secondly, SVBRDF and its variants, being independent of geometry, demand more samples to mitigate noise arising from complex occlusion within an object.

Precomputed Radiance Transfer (PRT) \cite{sloan2023precomputed}, addresses these issues by precomputing and integrating the material response and incident radiance across all possible incident light directions $\omega_i$ as transfer coefficients and light coefficients, respectively. Thus it is able to produce noise-free rendering results with sufficient samples in the precomputation stage. 
Moreover, the transfer coefficients are multiplied by a visibility term which encodes the intricate self-occlusion of objects, which simplifies the interaction between light and objects, facilitating noise-free rendering of complex geometries. 
The equation of PRT can be written as:
\begin{equation}
\label{eq:03}
    L_o(x_o,\omega_o) = \sum_{k=1}^K l_k(x_o,\omega_o) (t_{k_1}(x_o,\omega_o)+t_{k_2}(x_o,\omega_o)+...),
\end{equation}
where $K$ denotes the number of coefficients, $l_k$ represents the light coefficient for the $k$-th term, $t_{k_1}$ is the corresponding transfer coefficient accounting for the lights directly emitting from the light sources, and $t_{k_{2,3,...}}$ is the interreflected transfer coefficients accounting for indirect lights from the object surface itself bouncing for 2 or more times. 
Further breakdown of Equation (\ref{eq:03}) is provided in Equations (\ref{eq:04}) and (\ref{eq:05}):
\begin{equation}
\label{eq:04}
    l_k(x_o,\omega_o) = \int_{\Omega}L_i(\omega_i)\beta_k(\omega_i)d\omega_i,
\end{equation}
\begin{equation}
\label{eq:05}
    t_{k_1}(x_o,\omega_o) = \int_{\Omega}f(x_o,\omega_i,\omega_o)V(x_o, \omega_i,t)(\omega_i \cdot n)\beta_k(\omega_i)d\omega_i,
\end{equation}
where $\beta$ is an infinite set of basis functions used to project and reconstruct arbitrary functions, and $V(x_o,\omega_i,t)$ is a visibility term that equals 1 when the ray from $x_o$ to direction $\omega_i$ does not intersect the object itself, and 0 otherwise. 
% \textcolor{yijie}{Traditional PRT calculates interreflected transfer coefficients $t_{k_{2,3,...}}$ to account for inter-object indirect lighting. However, this calculation is computationally expensive and becomes invalid when objects move. To address these limitations, we propose transferring the responsibility of inter-object reflection to light probes. While traditional PRT requires interreflected transfer coefficients because its lighting coefficients are computed from a global environment map (which only accounts for one-bounce direct lighting), our approach incorporates multi-bounce lighting directly in near-object light probes so that inter-object lighting effects are stored in probe light coefficients. This strategy effectively compensates for inter-object lighting while supporting dynamic scenes with moving objects and changing lighting conditions.}
Following the derivation of $t_{k_1}$, the subsequent computation of interreflected transfer coefficients $t_{k_{2,3,...}}$ can be accomplished by referring to the methodology outlined in \cite{sloan2023precomputed}.
% However, due to the computationally intensive nature of this derivation process, it is omitted from the current implementation of this paper. 
% Instead, in Section \ref{subsec:compensation}, we will demonstrate that the representation of indirect light transport encapsulated in $t_{k_{2,3,...}}$ can be effectively substituted with our novel near-place probe-based light coefficients. Consequently, the calculation of $t_{k_{2,3,...}}$ becomes uncecessary.

However, PRT's conventional approach to material representation using 1-D vectors for diffuse materials and 2-D matrices for glossy materials \cite{sloan2023precomputed} limits its support to simple lighting models such as the Lambertian model and Phong/Blinn-Phong model, thus constraining the representational capacity of the SVBRDF function $f$ to only diffuse and glossy materials. Furthermore, PRT assumes distant lighting, further restricting its practical utility. Additionally, to accurately capture inter-object indirect lighting effects, PRT requires the calculation of interreflected transfer coefficients $t_{k_{2,3,...}}$, which are not only computationally expensive but also become invalid when objects move within the scene, making dynamic scenes particularly challenging.

To overcome these limitations, we propose an object-centric neural transfer model that extends the expressive power of materials in global illumination. This model utilizes vertex-attached latents and a neural transfer decoder to represent an object's intricate self-occlusion and a broad range of reflective materials, including the complex Disney BRDF model\cite{burley2012physically} and even measured realistic materials lacking an explicit analytic formulation. 
% \sout{To accommodate varying lighting conditions, including those in close proximity, we introduce a real-time light probe generation method with an across-probe radiance-sharing strategy to illuminate the scene effectively, these light probes considers inter-object lighting effects by shading multi-bounce light rays into the scene and convert to SH light coeffiencients, accounting for indirect lighting without calculating the complicated interreflected transfer coefficients.} 
To accommodate dynamic and varying lighting conditions, including light sources in close proximity, we introduce a real-time light probe generation method with an innovative across-probe radiance-sharing strategy. These light probes effectively capture inter-object lighting effects by tracing multi-bounce light rays throughout the scene and converting them to spherical harmonic light coefficients, thereby accounting for indirect lighting without the computational burden of calculating the complicated interreflected transfer coefficients.

\section{Object-Centric Neural Transfer Model}
\label{sec:appearance}
% \textcolor{red}{Title: Object-Centric Neural Appearance Model?}

% \textcolor{red}{[Better to use few sentences to summarizes the relationship between this section and this paper.]}
% \textcolor{red}{ [Unlike conventional real-time / offline rendering approach using simple analytical hand-crafted functions (E.g., Disney BRDF~\cite{}) to represent material properties of surfaces that each pixels need to be analyzed independently,  we present a novel approach to model the incoming light at object-scale, while using data-driven pre-computed radiance transfer techniques to represent the local details of objects, thus with less computations without compensation of quality. Firstly, we describe the representation of the object-centric neural appearance model in Section \ref{subsec:model} and the detailed process of collecting data and training the model in Section \ref{subsec:train}we describe the representation of the object-centric neural appearance model in Section \ref{subsec:model} and the detailed process of collecting data and training the model in Section \ref{subsec:train}]}

Unlike conventional rendering approaches that leverage analytical hand-crafted functions like the Disney BSDF \cite{burley2012physically} to represent materials, which can only respond to a single ray direction, we propose to represent full-range material responses in the frequency domain using a novel object-centric neural transfer model. Thus, with only one inference, the full-range responses of materials can be evaluated, demonstrating the potential for efficient rendering without the need for expensive sampling or ray tracing. We present our method in two parts, the representation of the object-centric neural transfer model is first introduced in Section \ref{subsec:model} and the detailed process of collecting data and training the model is described in Section \ref{subsec:train}

\subsection{Network Architecture and Representation}
\label{subsec:model}

Our aim with the transfer model is to directly generate high-fidelity, view- and position-dependent transfer coefficients $t$ derived from intricate reflective materials and self-occlusion within objects. 
In contrast to approaches like \cite{zeltner2023real}, which utilize neural networks to output BRDF values $f(x_o,\omega_o,\omega_i)$ for individual incident lighting directions $\omega_i$, our method for predicting transfer coefficients $t_k(x_o,\omega_o)$ yields noise-free rendering outcomes, as $t_k$ encompasses a comprehensive integral over $\omega_i$. 

Technically, our object-centric neural transfer model comprises two key components: \textit{vertex latents} and \textit{neural transfer decoder}. 
The \textit{vertex latents} are a set of $d$-dimensional vectors attached at every vertex of an object, encoding fundamental per-vertex transfer properties.
And the \textit{neural transfer decoder} is a multi-layer perceptron (MLP), which can be either object-specific or scene-specific. 
The object-specific decoder encodes transfer properties for a single object, offering compactness and precision. 
However, instantaneously and concurrently inferring thousands of separate neural networks poses challenges for real-time implementation. 
Therefore, the default choice for this paper is the scene-specific decoder, trained to generalize across all objects within the scene. 
This decoder can accommodate rigid object movements and rotations within the scene and has demonstrated its ability to produce high-quality rendering results in experiments.

A fully-trained model facilitates the generation of view- and position-dependent transfer coefficients:
\begin{equation}
\label{eq:06}
    t_{0-K}(x_o,\omega_o) = \Phi(z(x_o),\omega_o,n,\alpha),
\end{equation}
where $\Phi$ is the neural transfer decoder, $\alpha$ denotes a three-channel surface albedo serving as supplementary input enhancement, the surface shading normal $n$ provides additional geometric information, and $z(x_o)$ denotes the barycentric-interpolated latent at surface position $x_o$:
\begin{equation}
\label{eq:07}
    z(x_0) = \lambda_1z(i_1)+\lambda_2z(i_2)+\lambda_3z(i_3),
\end{equation}
where $\lambda_{1-3}$ are the barycentric coordinates that sum up to 1, and $i_{1-3}$ represent the vertex indices of the triangle in which the surface position $x_o$ is situated.

\subsection{Data-driven modeling}
\label{subsec:train}
We collect the training data by uniformly sampling surface positions of an object and outgoing directions. Each complete sample set comprises the input part $\{i_{1-3},\lambda_{1-3},x_o,\omega_o,n,\alpha\}$ and the output part $t_{0-K}(x_o,\omega_o)$ for the training process.

To generate the input part, we first select a triangle based on its ratio of area relative to the entire surface area of the object.
$i_{1-3}$ denote the three vertex indices of the chosen triangle. Subsequently, we randomly generate a set of barycentric coordinates $\lambda_{1-3}$ to determine a random position $x_o$ within the triangle. 
The albedo $\alpha$ is obtained by querying the material of the triangle, while the surface shading normal $n$ is interpolated using the barycentric coordinates and the corresponding normal directions at vertices $i_{1-3}$. 

Concerning the output part, as inferred from Equation (\ref{eq:05}), it involves an integral over $\omega_i$, so for each sample, we generate numerous directions of $\omega_i$ to achieve a comprehensive integration, typically set to 2000 in our default implementation.
The BRDF $f(x_o,\omega_i,\omega_o)$ can be any complex SVBRDF model with spatially varying roughness, metallicity, anisotropic response, and other properties challenging to describe with analytical formulas.
The visibility term $V(x_o,\omega_i)$ needs to be modulated by a distance threshold $t$, denoted as $V(x_o,\omega_i, t)$. This term evaluates to 1 when the ray from $x_o$ to direction $\omega_i$ does not intersect the object or intersects but the distance between $x_o$ and the intersection point exceeds $t$; otherwise, it is 0. This modification is necessary because the original PRT assumes distant light sources. However, our method aims to accommodate nearby variant lights. If a light source exists within an object, the original visibility term could yield erroneous rendering results, surfaces within the object will be entirely dark.
There are several options for the basis functions $\beta$, including wavelets, spherical Gaussians, and spherical harmonics. We opt for spherical harmonics due to their advantageous properties such as orthogonality and rotation invariance. 
Therefore, a generated sample of $t_0(x_o,\omega_o)$ can be written as:
\begin{equation}
\label{eq:08}
    t(x_o,\omega_o) = \sum_{l=0}^{L}\sum_{m=-l}^{l}
    \int_{\Omega}f(x_o,\omega_i,\omega_o)V(x_o, \omega_i,t)(\omega_i \cdot n)Y_l^m(\omega_i)d\omega_i,
\end{equation}
where $Y_l^m$ is a spherical harmonic function of degree $l$ and order $m$, $L$ denotes the maximum degree, and the number of coefficients $K$ equals to $(L+1)^2$. 

The training process is straightforward, entailing the joint optimization of vertex latents and the neural transfer decoder. 
In practice, we assemble a batch of training samples to construct $z(x_o)$ from $\lambda_{1-3}, i_{1-3}$ and randomly initialized vertex latents $z$.
Subsequently, $z(x_o)$ is concatenated with $\omega_o,n,\alpha$ to form the complete input channels, which are then fed into the neural transfer decoder to generate the predicted transfer coefficients. Both components are optimized using L1-loss, comparing the predicted and ground truth transfer coefficients.

Upon training the model for the initial transfer coefficients $t_0$, datasets containing transfer coefficients for additional inter-object light bounces can be derived using the trained model, following the approach outlined in \cite{sloan2023precomputed}. This process enables the neural transfer model and the vertex-attached latent features to be further trained, consequently encoding the intricate properties of inter-object reflections within the learned representations.

\begin{figure}
    \centering
    \includegraphics[width=1\linewidth]{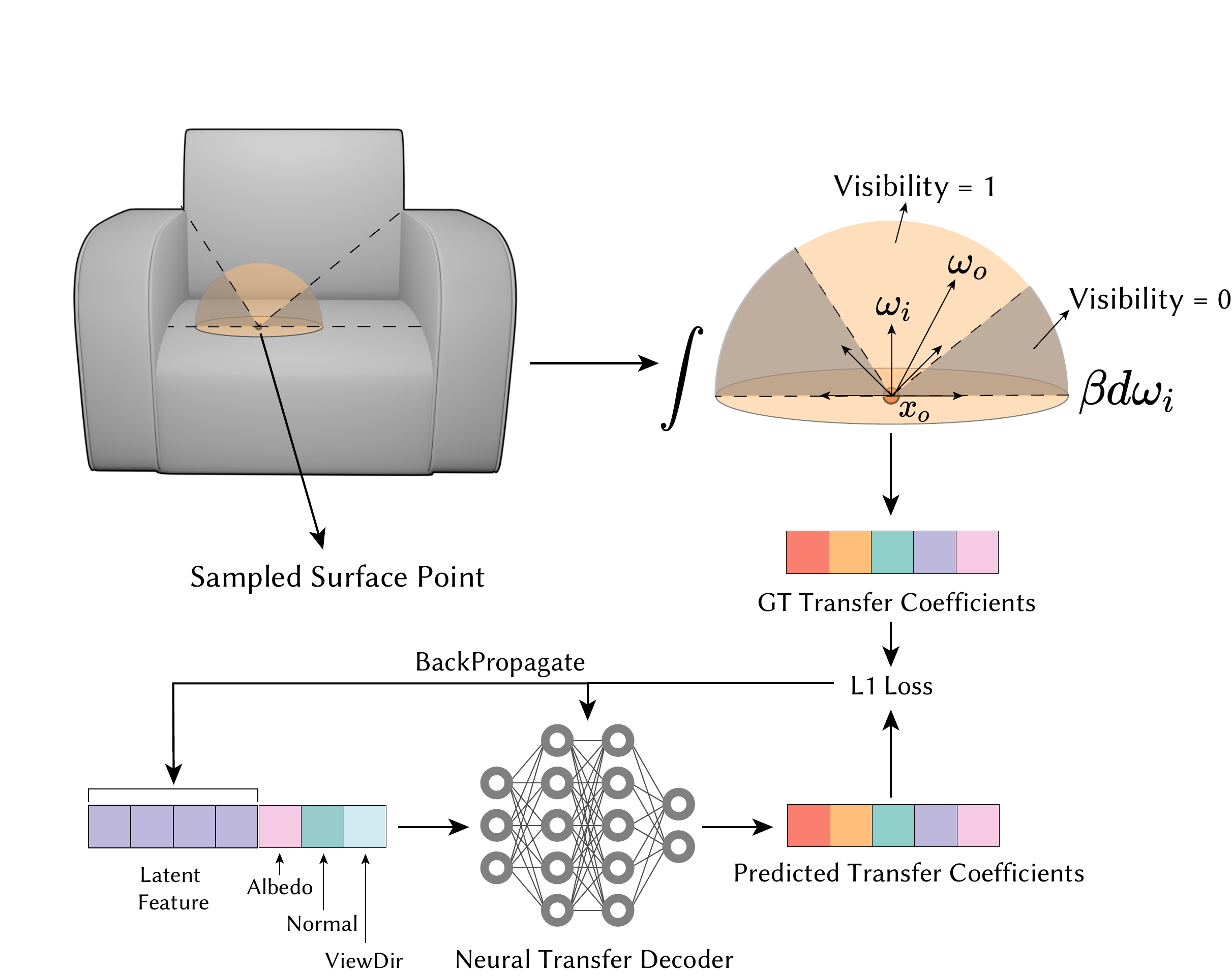}
    \caption{Data collection and training process for our appearance model. Top row: A random surface point $x_o$ is sampled together with the outgoing direction $\omega_o$, and valid material evaluations (that do not intersect with the surface itself) are integrated over the incident direction $\omega_i$, followed by multiplying the basis function $\beta$ to get the ground truth (GT) transfer coefficients. Bottom row: The interpolated latent features are concatenated with necessary G-buffers and then fed into the neural transfer decoder to output the predicted transfer coefficients. The latent features and the neural transfer decoder are jointly optimized by backpropagating the L1-loss.}
    \label{fig:collectData}
\end{figure}

The data collection and training process are illustrated in Figure \ref{fig:collectData}.

\section{Light Probe}
\label{sec:probe}

In this section, we present our real-time lighting system, constructed from sparsely scattered light probes within the scene. We explain the representation of these light probes in Section \ref{subsec:probe_coe}, while Section \ref{subsec:radiance_sharing} outlines our methodology for generating real-time light probes through an innovative across-probe radiance-sharing strategy. Finally, we introduce the interpolation strategy of light probes in Section \ref{subsec:probe_interpolation}.

\subsection{SH Light Probes}
\label{subsec:probe_coe}

As we adopt spherical harmonics as the basis functions to represent transfer coefficients, the light coefficients shown in Equation \ref{eq:04} should also be projected and represented using spherical harmonics:
\begin{equation}
\label{eq:09}
    l_k(x_o,\omega_o)=\sum_{l=0}^{L}\sum_{m=-l}^{l}\int_{\Omega}L_i(\omega_i)Y_l^m(\omega_i)d\omega_i.
\end{equation}

Since global illumination mainly display the low-to-mid frequency lighting effects, it can be rendered smoothly with sparsely scattered light probes in the scene\cite{greger1998irradiance,mcguire2017real,silvennoinen2017real,majercik2019dynamic,datta2023adaptive}. 
We align with the convention of storing radiance in sparsely-placed probes within the scene. However, diverging from the practice of storing a radiance map in each probe, we store a set of light coefficients generated by projecting the radiance map onto the spherical harmonic basis in the frequency domain. We refer to our probes as \textit{SH Light Probes}.

To account for inter-object lighting effects, we allow the incoming light $L_i(\omega_i)$ to bounce multiple times in the scene. Specifically, in the probe-based lighting system, we perform multi-bounce ray tracing for each probe to compute the pixel colors of the radiance map. This approach captures inter-object lighting interactions and eliminates the need to calculate the inter-reflected transfer coefficients $t_{k_{2,3,...}}$ in Equation \ref{eq:03}. Consequently, we can approximate Equation \ref{eq:03} as follows:

\begin{equation}
\label{eq:added}
    L_o(x_o,\omega_o) = \sum_{k=1}^K (l_{k_1}(x_o,\omega_o) + l_{k_2}(x_o,\omega_o) + \ldots)t_{k_1}(x_o,\omega_o),
\end{equation}
where $l_{k_1}(x_o,\omega_o)$ represents the light coefficients for direct incoming light, and $l_{k_{2,3,...}}(x_o,\omega_o)$ represents the light coefficients for indirect incoming light generated by multiple ray bounces in the scene. This transformation can be computed efficiently with modern rendering engines such as Falcor\cite{kallweit22falcor}, which support multi-bounce ray tracing at minimal computational cost. This approach enables effective lighting of movable objects since inter-object lighting is handled by real-time generated light probes rather than being encoded in the transfer coefficients.

%Notably, \cite{sloan2023precomputed} elucidates that $L_i(\omega_i)$ represents the incident radiance for an object, assuming the object's removal from the scene. 
%This notion implies that in a scene comprising thousands of objects, the probes for each object must be generated independently and utilized solely by that object, as the color of the same ray might be different once an object is removed. 
%While theoretically sound, 
%However, in practice, we place probes at scene scale rather than object scale considering the following two primary reasons:
%\begin{itemize}
    %\item The low-to-mid frequency global illumination exhibits minimal sensitivity to the removal of a single object, resulting in a negligible impact on the scene radiance.
    %\item Object-scale probes would engender substantial probe redundancy, as in practical implementations, the incoming radiance at each shading point is interpolated using the nearest 4/8 probes. Consequently, even a minuscule object would necessitate at least 4 probes, which are not shared by its other nearest tiny objects. This would introduce significant computational overhead.
%\end{itemize}

%Therefore, in our default implementation, we position $N \times N \times N$ SH light probes within the scene to encompass the scene bounds.

\subsection{Across-probe Radiance Sharing}
\label{subsec:radiance_sharing}

Regarding the generation of a single light probe, traditional approaches tend to employ path-tracing to compute the pixel color of the irradiance map. However, this method inevitably leads to a scaling problem, wherein as the resolution of the irradiance map and the number of light probes increase, the computational overhead for path-tracing will also escalate.

To address this, we propose the across-probe radiance-sharing strategy, aiming to decouple the linear relationship between computation and probe parameters. The main insight of this strategy is that the low-to-mid frequency per-ray radiance can be shared by different light probes. This strategy involves two steps: \textit{Per-Probe Radiance Scattering} and \textit{Atomic Probe Map Rasterization}.

\textit{Per-Probe Radiance Scattering.} Suppose we place $N\times N\times N$ \textit{SH Light Probes} in the scene. For each probe, $M$ rays are cast from the probe's location in random directions, excluding rays that directly intersect light sources, as only global illumination is accounted for. For each frame, the shaded points are rendered with a single path sample, and the noise caused by insufficient samples is mitigated through pixel accumulation during the rasterization stage. Consequently, after each frame, we obtain a total of $M \times N^3$ points, each associated with a radiance color and a world-space position.

\textit{Atomic Probe Map Rasterization.} Once we obtain a set of shaded points, we can perform point cloud rasterization to generate the irradiance map for each probe. However, naively rasterizing all points to each probe is not temporally efficient, as the point-to-probe rasterization operation cannot be effectively parallelized due to potential conflicts when multiple points attempt to write to the same pixel, causing memory synchronization issues. Additionally, there exists a scalability problem wherein rasterizing all points to each probe would lead to $M \times N^3 \times N^3$ rasterization operations. To mitigate these issues, we introduce two modifications. First, we treat each rasterization operation as an atomic operation, akin to \cite{ras1,ras2}, which leverages shader atomic operations that encompass uninterrupted read and write operations to ensure correct results. This approach enables effective parallelization of the point-to-probe rasterization operation on the GPU, enhancing efficiency. Second, for each probe, we rasterize its irradiance map using its own shaded points and the shaded points from its 3D nearest 26 probes, instead of utilizing all points. By doing so, we increase the sample rate for a single probe and reduce the number of rasterization operations from $M \times N^3 \times N^3$ to $M \times 27 \times N^3$. We show a 2D example of this rasterization process in the bottom left part of Figure \ref{fig:pipeline}.

Our strategy lessons the scalability issue dramatically by assuming a relatively small number of per-probe rays $M$ in conjunction with the rasterization technique. Traditionally, the color of each pixel on an irradiance map is independently path-traced, necessitating an increase in the number of path-tracing operations to improve the probe quality, i.e., the map resolution. However, as the per-ray radiance is shared across multiple probes in our strategy, the resolution of the probe irradiance map is decoupled from the number of per-probe path-tracing operations $M$. This affords us the opportunity to generate high-resolution probe irradiance maps with only a small value of $M$ through rasterization. Moreover, our atomic probe map rasterization technique can be performed efficiently in real-time implementations, enabling rapid mitigation of noise in the irradiance maps by accumulating the rasterized points at each frame. 
% A qualitative comparison between the irradiance maps generated by our rasterization technique and independent path-tracing with 100 light samples is presented in Figure \ref{fig:map_comparision}.
% [Zhi Liang Bu Hao, Bu Hao Bi Jiao]

\subsection{Probe Interpolation}
\label{subsec:probe_interpolation}

Since the light probes are scattered in a regular grid, we use trilinear interpolation of the 8 nearest probes to estimate the incoming radiance at a shading point. To minimize light/shadow leak artifacts, we employ similar pruning techniques as in DDGI\cite{majercik2019dynamic}, which involves discarding the probes that are occluded by other surfaces and the probes that are not on the same side of the shading point with respect to the surface geometry. If all 8 probes are pruned, we perform a weighted average of their contributions based on their distance to the shading point and the cosine angular weight between the shading point normal and the vector towards each probe, which is similar to the approach used in DDGI\cite{majercik2019dynamic}.

\section{Experiments}
\label{sec:experiments}

\begin{figure*}
    \centering
    \includegraphics[width=1\linewidth]{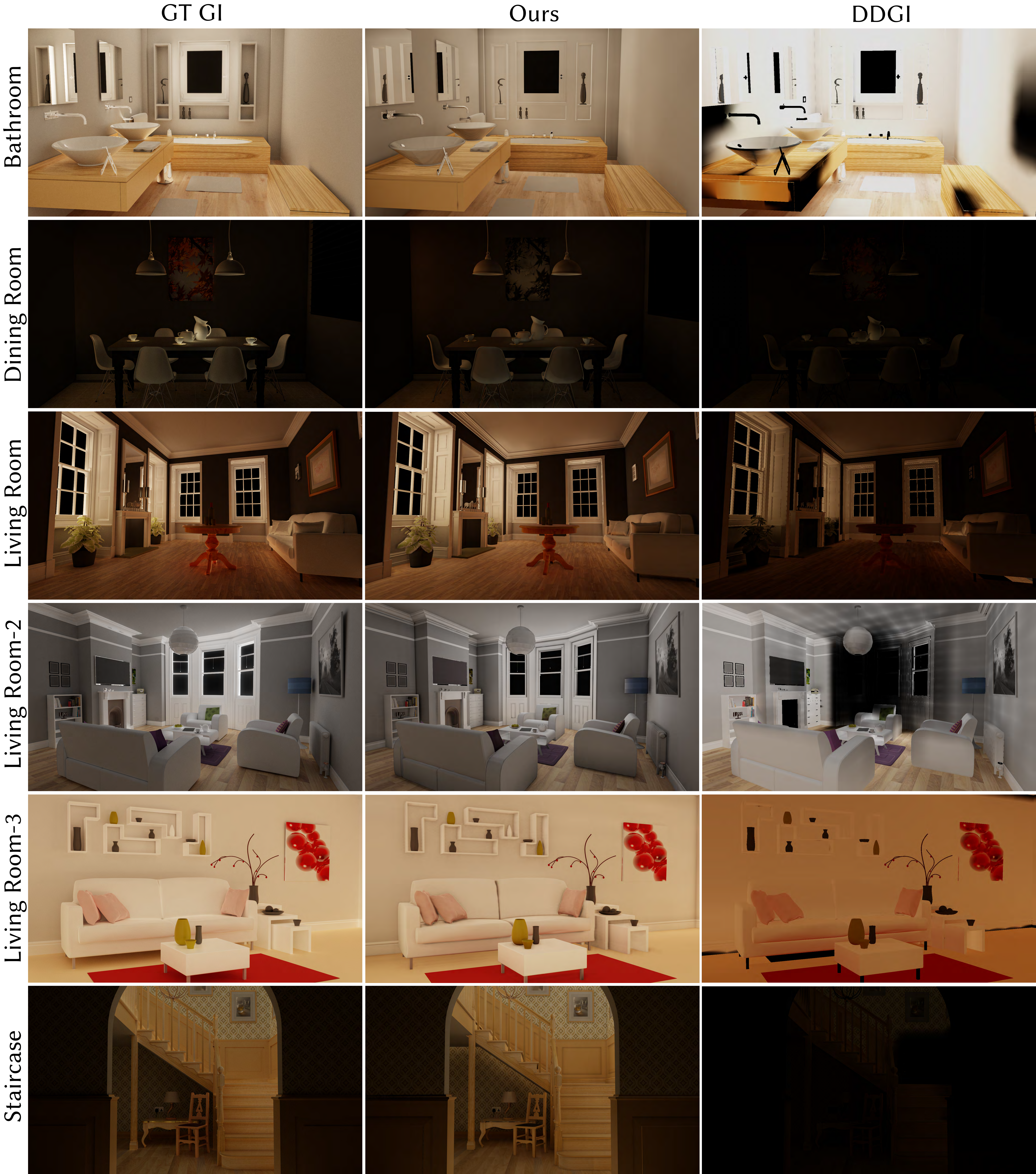}
    \caption{Qualitative comparison in terms of global illumination on the selected 6 scenes from the Bitterli dataset \cite{resources16}.}
    \label{fig:gi_comparison}
\end{figure*}

\begin{figure*}
    \centering
    \includegraphics[width=1\linewidth]{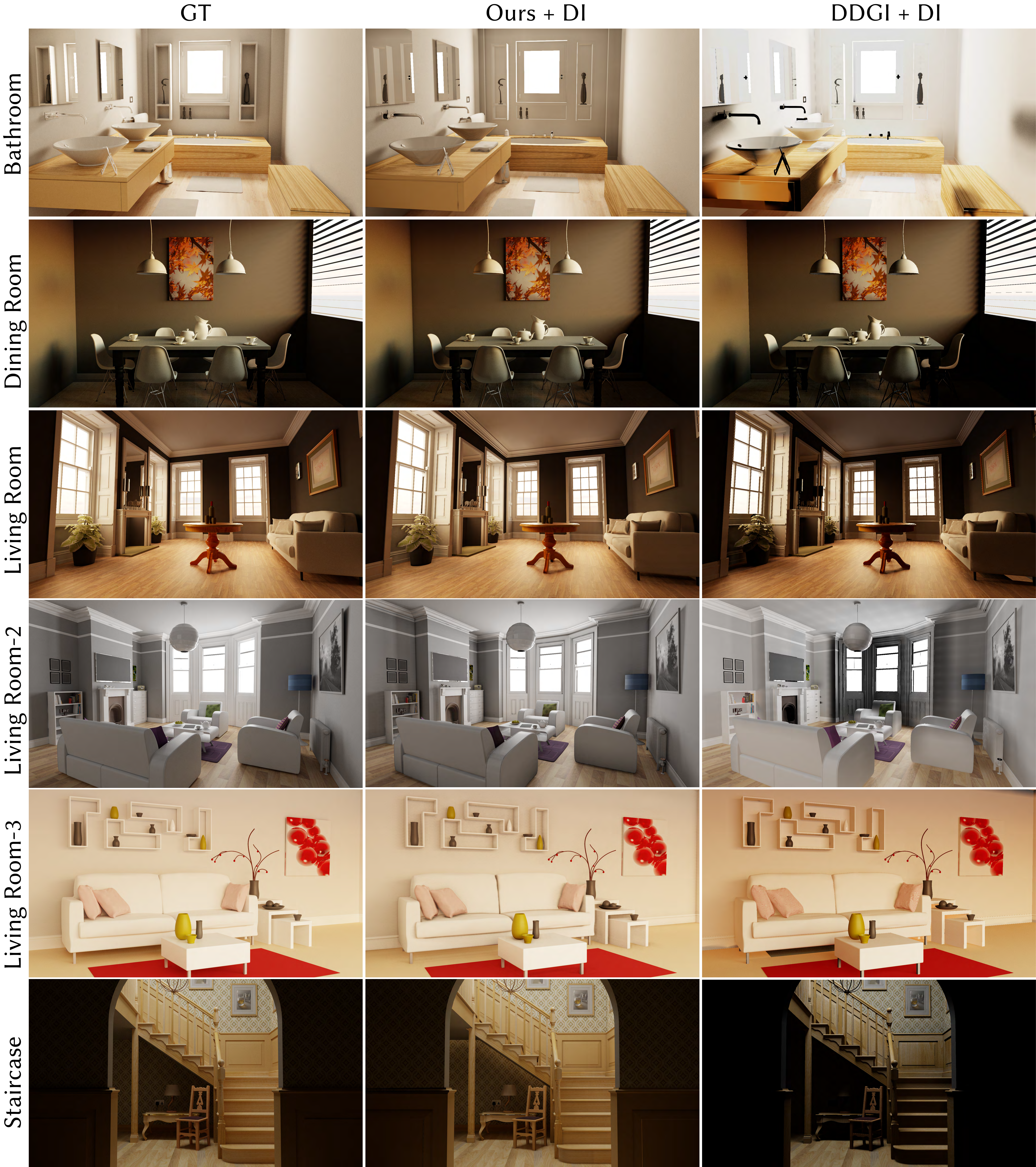}
    \caption{Qualitative comparison in terms of global illumination+direct illumination on the selected 6 scenes from the Bitterli dataset \cite{resources16}.}
    \label{fig:gt_comparison}
\end{figure*}

\subsection{Experiments Setup}
\label{subsec:setup}

\textit{Sample Scenes.} We evaluate the scene-scale global illumination on several modified scenes from the Bitterli dataset \cite{resources16}. All scenes are modified to exclude transparent objects but can encompass multiple light sources and hundreds of reflective objects with materials exhibiting diverse effects, such as diffuse, coated diffuse, rough plastic, metallic, and conductive surfaces. The lighting configurations in these scenes include area lights for individual windows, local point light sources for small fixtures such as light bulbs, and distant environment maps for certain open scenes. In addition, we evaluate three object-scale scenes from the \textit{Grand Bazaar}\footnote{https://www.bigmediumsmall.com/grand-bazaar}, where each scene contains either a metallic object or a complex-textured object, illuminated solely by an environment map. These scenes are used to assess the ability of our neural transfer model to represent complex, multi-layered material effects.

\textit{Baselines.} We evaluate the methods under two setups: global illumination (GI) and global illumination combined with direct illumination (GI+DI). For the GI setup, we compare our method against Dynamic Diffuse Global Illumination (DDGI)\cite{majercik2019dynamic} and path-traced results without direct illumination. For the GI+DI setup, we incorporate the path-traced direct illumination component into both our method and DDGI for fair comparison with the path-traced reference image. 
% For practical use, we can choose some real-time direct illumination methods such as the instant denoised direct illumination method proposed by NVIDIA\footnote{https://developer.nvidia.com/rtx/ray-tracing/rt-denoisers}, however, such methods will introduce extra inaccuracy. 
Subsequently, we compare our results with DDGI+DI and path-traced results encompassing both global and direct illumination.

We also compare our method against NRC\cite{10.1145/3450626.3459812}, a global illumination technique that uses a single MLP to cache radiance. We adopt its open-source implementation from RTXGI\cite{RTXGI}. However, because RTXGI and Falcor use different scene formats, \textit{Living Room-2} is the only scene they both support. In addition, their distinct code implementations introduce some unavoidable rendering differences. Consequently, we compare our method and NRC only qualitatively in Section \ref{sec:qualitative}.

\textit{Configuration.} To evaluate the scene-scale illumination, for both our method and DDGI\cite{majercik2019dynamic}, a grid of $8\times 8\times 8$ light probes are placed in each scene. Each probe in DDGI shoots 256 rays per frame, while each probe shoots 100 rays in our method. The path-traced referenced images are generated by accumulating 20000 samples per pixel. As for object-scale illumination, the only light source is an environment map which is projected onto the spherical harmonics basis to get a set of light coefficients. And the default maximum degree of spherical harmonics is set to 4, balancing its ability of representation and efficiency.

\textit{Metrics.} We evaluate the performance of the methods in terms of both computational efficiency and rendering quality. For temporal efficiency, we evaluate our training efficiency on object-scale scenes to show our fast convergence. And for inference efficiency, we compare all methods using the average time required to render a single frame. As for rendering quality, we assess the methods by computing three quality metrics with respect to the path-traced reference image: the root mean squared error (RMSE), which measures the average magnitude of the errors; the L1 loss, which captures the average absolute difference; and the structural similarity (SSIM) index, which evaluates the perceptual similarity by considering luminance, contrast and structural factors.

\textit{Implementation.} To eliminate potential influences arising from disparities in rendering engines, shading languages, and experimental hardwares, we implement our method in the real-time rendering engine Falcor \cite{kallweit22falcor} and also execute DDGI and path-tracing within the same framework for fair comparisons. For the neural network components, we integrate the CUDA-based MLP framework tiny-cuda-nn \cite{tiny-cuda-nn} with Falcor, enabling seamless data transfer between the shader and CUDA environments. All experiments are conducted on a desktop computer equipped with an RTX 5000 GPU. For the scale of the training data of a scene, we sample 40,960,000 points per scene, where these sampled points are scattered on objects' surfaces weighted by their surface areas. And this sampling procedure can be done very efficiently using Falcor within 30 minutes. For training the neural transfer model, we employ a Fullyfused MLP from tiny-cuda-nn\cite{tiny-cuda-nn} with 3 hidden layers (128 dimensions each). The network takes 16-dimensional inputs (7D vertex features concatenated with 9D GBuffer data including albedo, normal, and view direction) and outputs 75-dimensional vectors corresponding to 4th-degree spherical harmonics. We optimize using Adam with an initial learning rate of 1e-4 and a MultiStepLR scheduler. With 2048 pixel samples per batch, training converges after approximately 200,000 iterations (~2 hours) while occupying less than 2 GB of video memory.

% \textcolor{red}{Overheads on memory}
% \textcolor{red}{Efficiency}

\subsection{Evaluations}
\subsubsection{Qualitative Evaluation}
\label{sec:qualitative}

To evaluate the scene-scale global illumination quality, we first present an overall comparative evaluation in Figure \ref{fig:gi_comparison} for global illumination (GI) and in Figure \ref{fig:gt_comparison} for global illumination combined with direct illumination (GI+DI). From an overall perspective, our method can provide more stable and visually plausible rendering results that are closer to the ground truth reference, compared to the DDGI method\cite{majercik2019dynamic}. Detailed qualitative comparisons and analysis are provided below:

\begin{figure}
    \centering
    \includegraphics[width=1\linewidth]{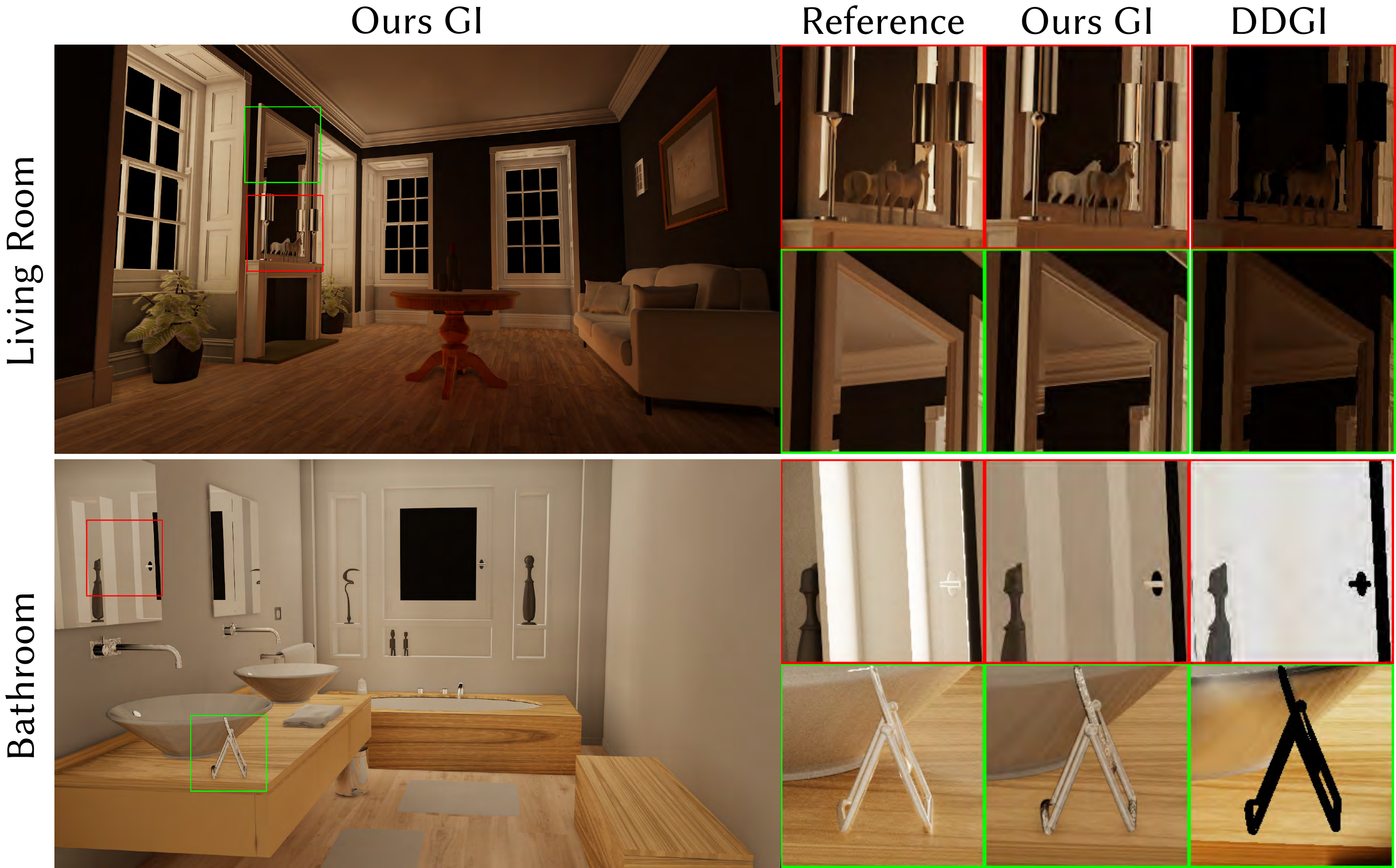}
    \caption{Qualitative GI comparison on specular/glossy effects.}
    \label{fig:mirror}
\end{figure}

\begin{figure}
    \centering
    \includegraphics[width=1\linewidth]{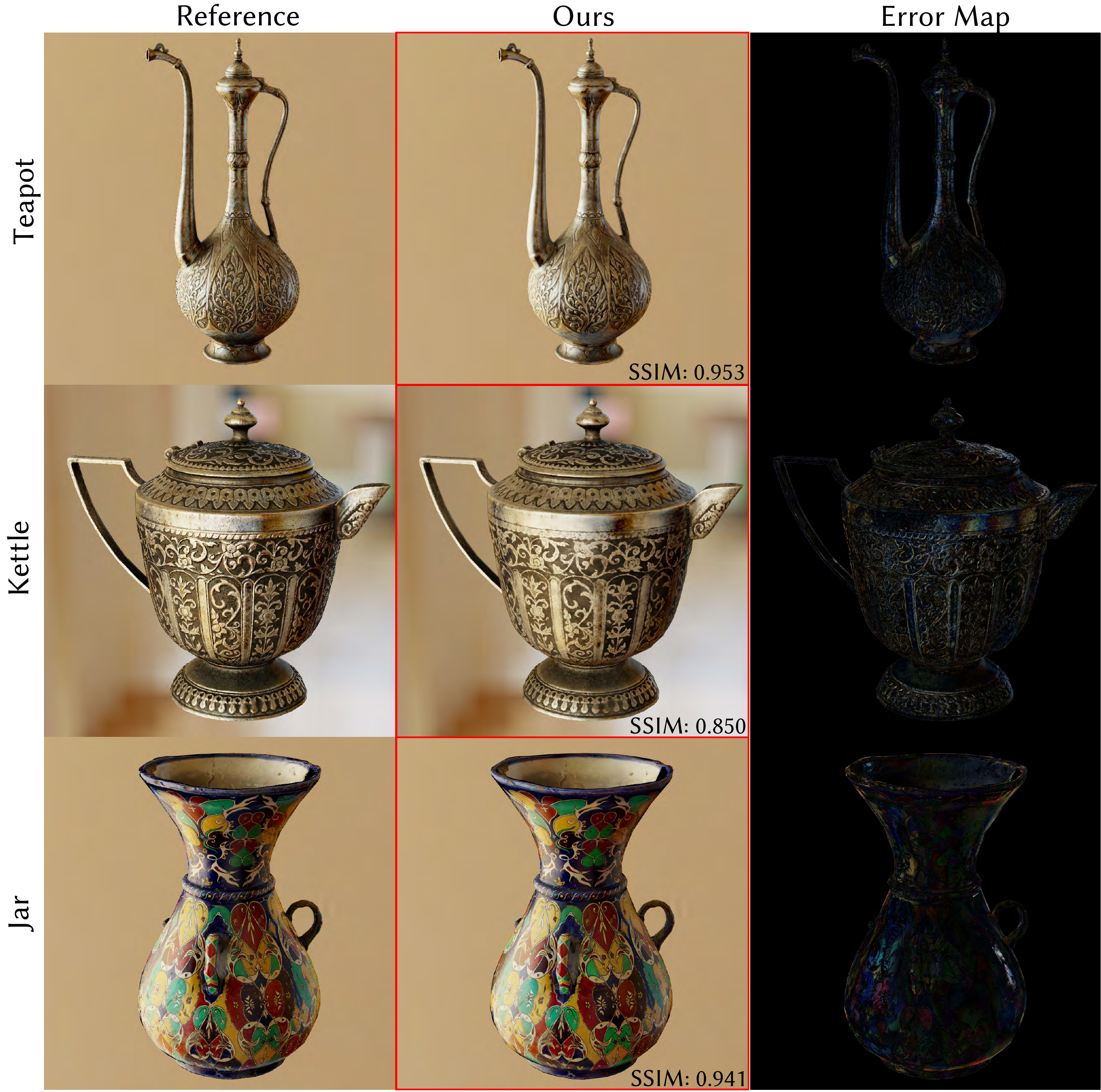}
    \caption{Object-level qualitative/quantitative comparison and error maps between our neural appearance model and the path-traced reference results.}
    \label{fig:single}
\end{figure}

\textit{Specular/Glossy Effects.} From Figure \ref{fig:mirror}, we can observe that our method produces reasonable glossy reflection effects on the lamps of the \textit{Living Room} scene and the mirror holder of the \textit{Bathroom} scene. However, as DDGI\cite{majercik2019dynamic} only accounts for diffuse materials, it cannot reproduce glossy effects. Furthermore, for the mirrors in both scenes, our method renders contents that are more accurate and visually closer to the ground truth reference compared to DDGI. Since our overall rendering result is closer to the reference image, the contents reflected in the mirrors, which involve one additional light bounce, are also more faithful to the reference. And since the scenes displayed in Figure \ref{fig:mirror} are overly complex with many other light transport effects, we especially show how our neural transfer model could represent the glossy reflection at object-level in Figure \ref{fig:single}. This figure specifically compares our method against path tracing reference renderings for challenging objects with complex multi-layered meterials, including metal teapots, a kettle, and a glossy jar. The results demonstrate that our neural transfer model successfully captures the nuanced characteristics of glossy materials and reproduces reflections that closely align with path-traced ground truth.

\begin{figure}
    \centering
    \includegraphics[width=1\linewidth]{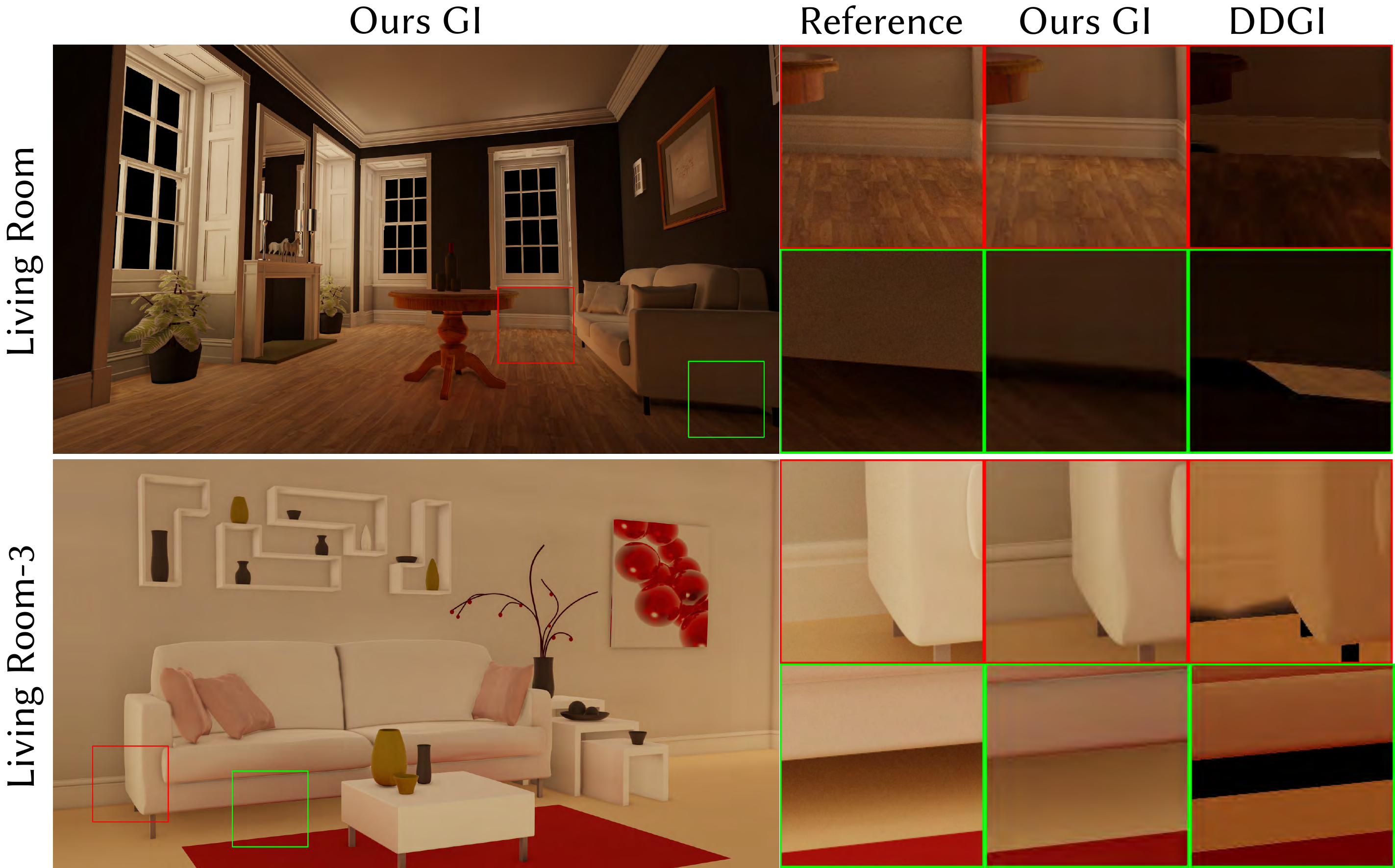}
    \caption{Qualitative GI comparison on light/shadow leaks.}
    \label{fig:leak}
\end{figure}

\textit{Light/Shadow Leak.} From Figure \ref{fig:leak}, we can observe that our method provides smoother rendering results without abrupt light or shadow leakage artifacts caused by probe interpolation. However, such leakage artifacts are evident at the edges of the rooms in the results produced by DDGI\cite{majercik2019dynamic}. This is because during probe interpolation, we strictly prune probes that are not on the same side of the surface geometry as the shading point, which effectively prevents light and shadow leaks for indoor environments. In contrast, DDGI takes all probes into account during interpolation, which may introduce light and shadow leakage artifacts.

\begin{figure}
    \centering
    \includegraphics[width=1\linewidth]{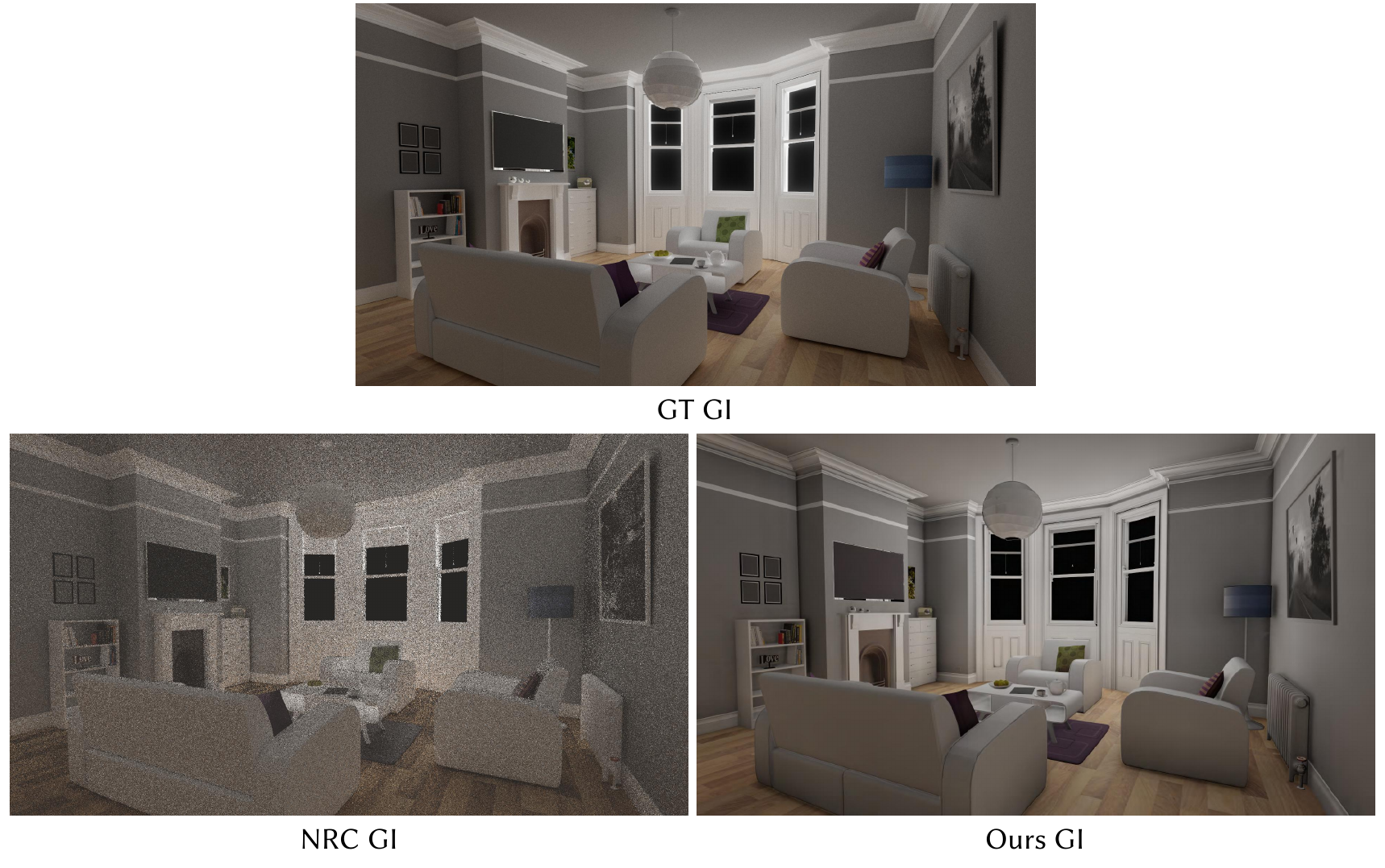}
    \caption{Qualitative comparison in terms of global illumination between our method and NRC\cite{10.1145/3450626.3459812} on the scene \textit{Living Room-2}.}
    \label{fig:nrc}
\end{figure}

We also compare our method with NRC\cite{10.1145/3450626.3459812} in Figure \ref{fig:nrc}. We can see that the NRC rendering suffers from visible noise. This is because NRC relies on a single global neural network to approximate the scene’s radiance at run time, which struggles to capture noise-free lighting details with limited samples.

\subsubsection{Quantitative Evaluation}

We provide quantitative comparisons for global illumination (GI) in Table \ref{tab:gi_comparison} and for the combined global illumination and direct illumination (GI+DI) in Table \ref{tab:gt_comparison}. The results demonstrate that our method achieves superior quality across all three evaluation metrics on these six scenes.

\begin{figure}
    \centering
    \includegraphics[width=1\linewidth]{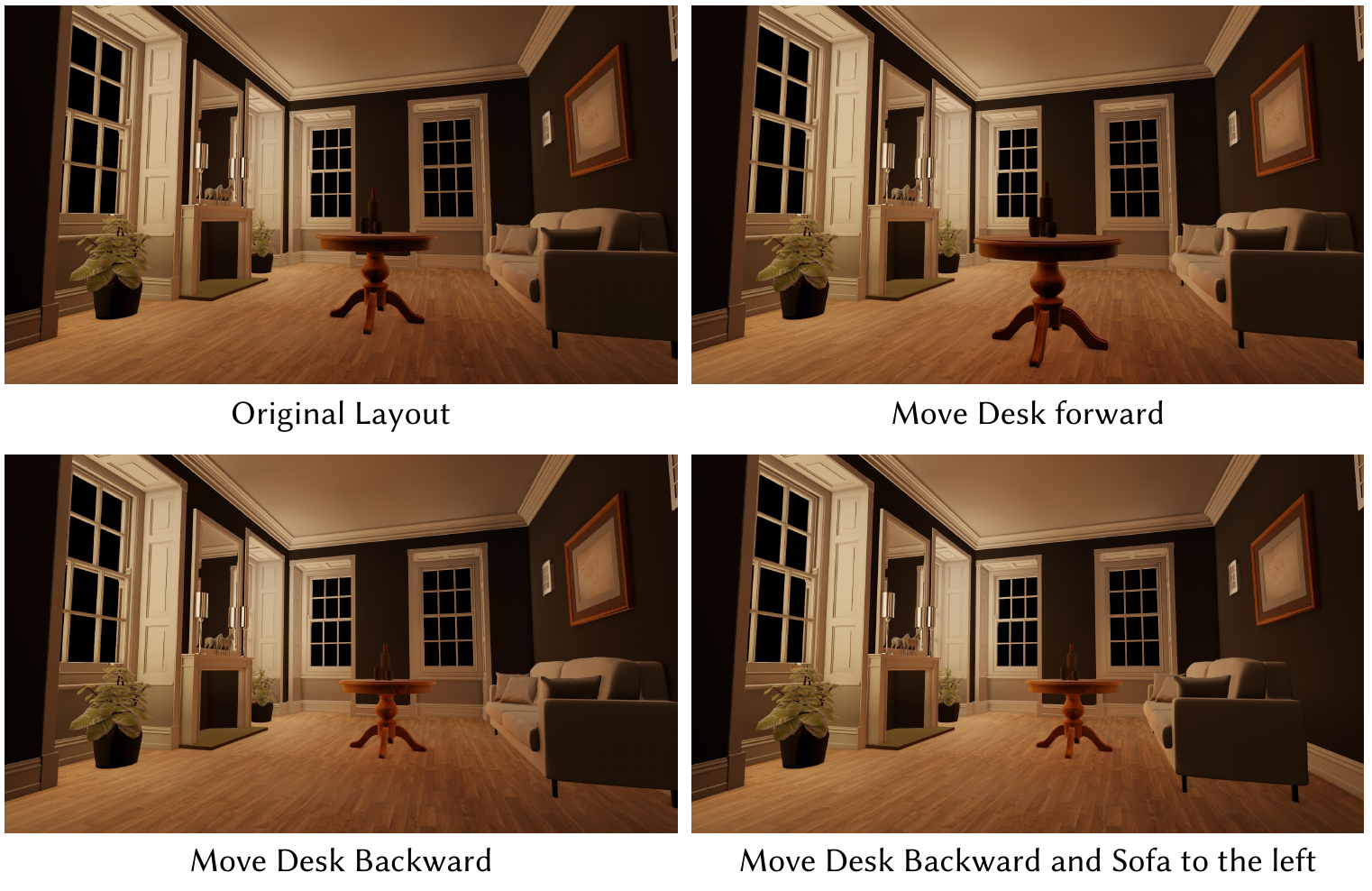}
    \caption{Visualizing global illumination effects in the \textit{Living Room} scene under various object displacement scenarios.}
    \label{fig:move}
\end{figure}

\subsubsection{Dynamic Movement Support}

Figure \ref{fig:move} presents global illumination renderings of the living-room scene with various object repositioning configurations. We demonstrate the robustness of our approach by testing two distinct displacement scenarios: translating the desk forward and backward along its axis, and shifting the sofa laterally to the left. The results confirm that our method maintains consistent and physically plausible lighting effects despite these object manipulations. This spatial invariance stems from our technique of embedding transfer features directly onto object vertices, enabling our approach to accommodate any rigid body transformation without retraining the scene network. Furthermore, as objects move, the spherical harmonic probes in the scene are dynamically updated, causing the light coefficients to adjust in real-time and produce visually coherent illumination results.

\begin{figure}[t]
    \centering
    \includegraphics[width=1\linewidth]{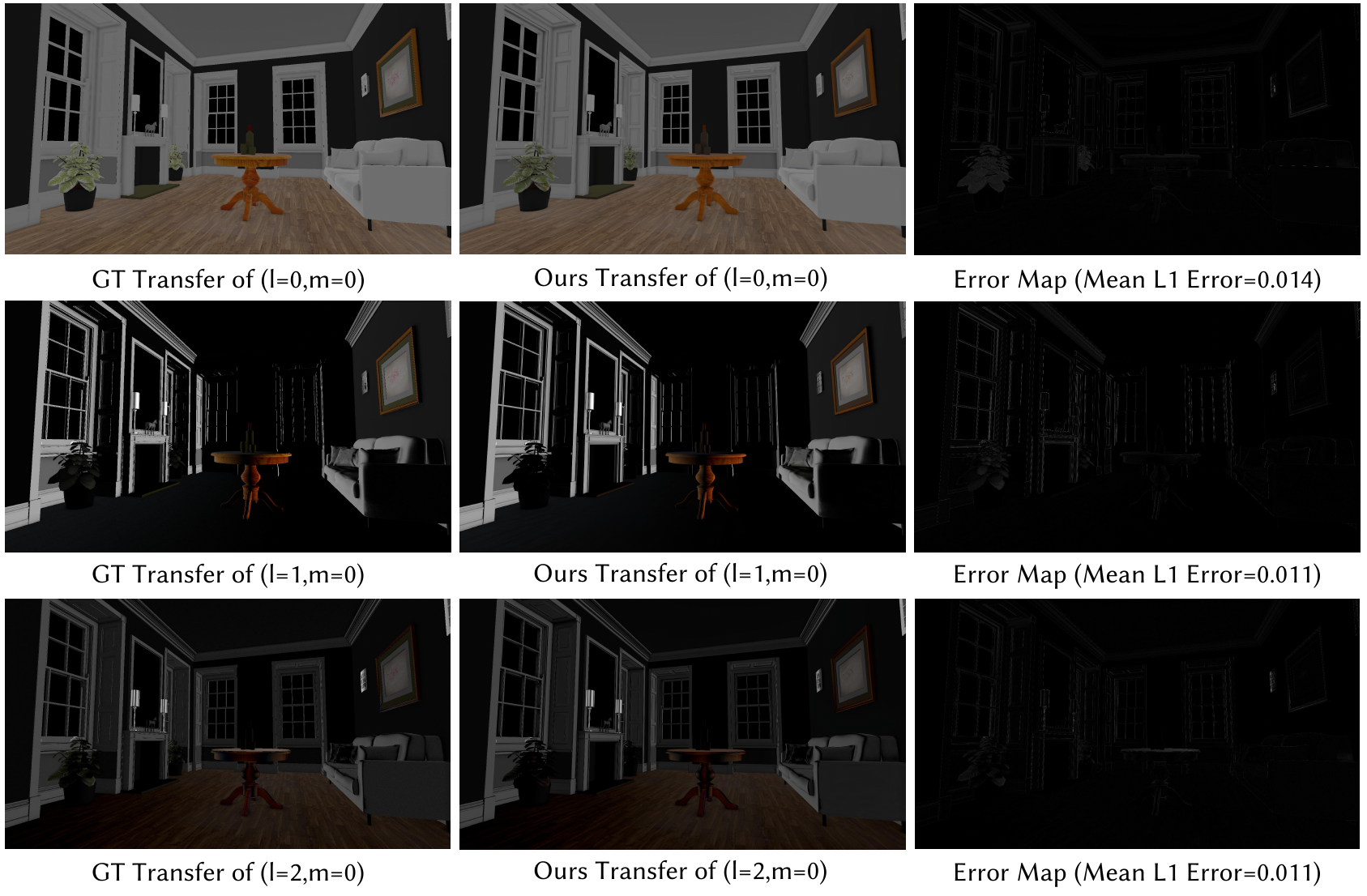}
    \caption{Qualitative/quantitative comparison of GT transfer coefficients and our predicted transfer coefficients on the scene \textit{Living Room}.}
    \label{fig:sh_compare}
\end{figure}

\begin{figure}[t]
    \centering
    \includegraphics[width=1\linewidth]{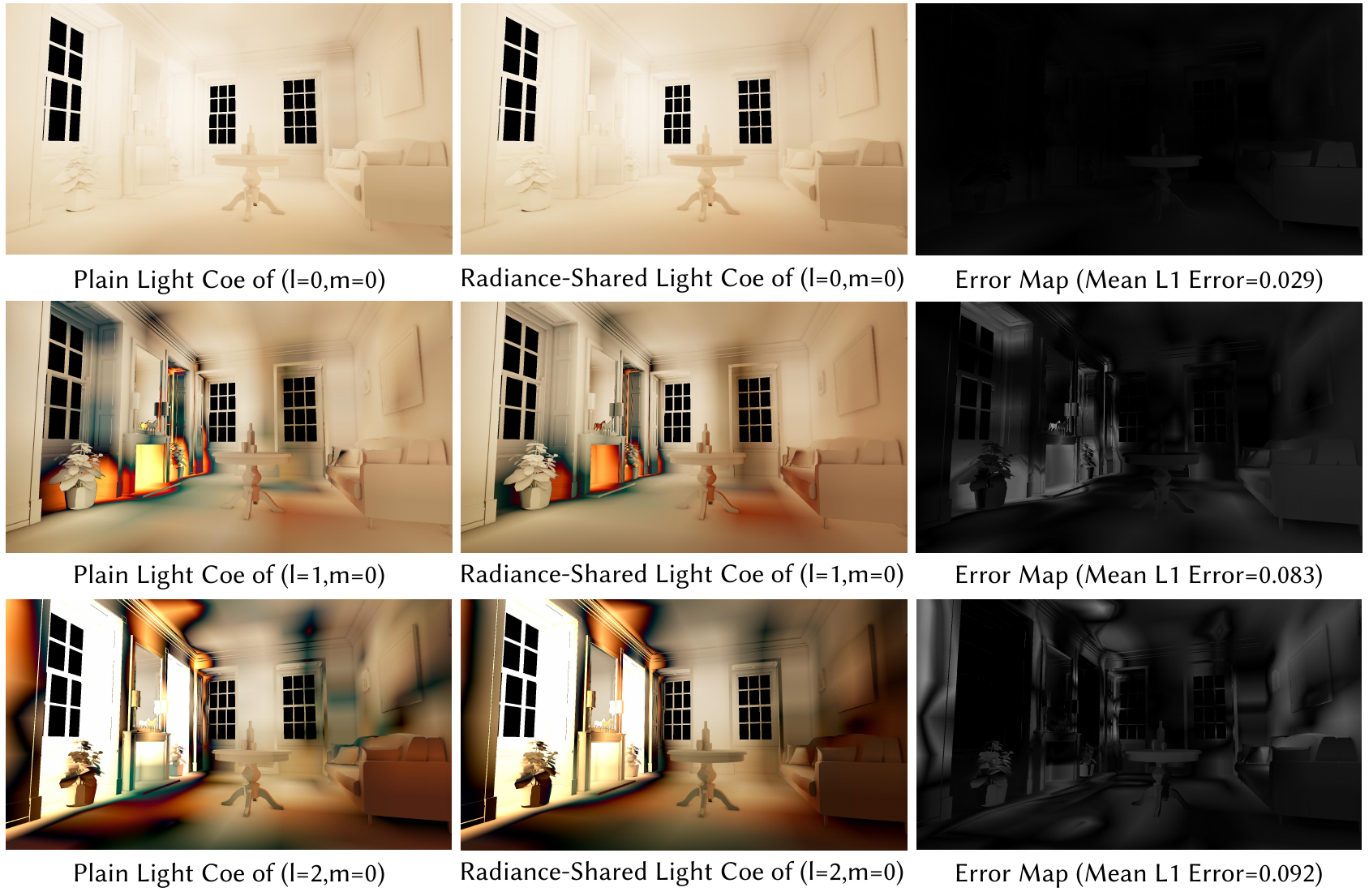}
    \caption{Qualitative/quantitative comparison of light coefficients interpolated using plain probe generation and across-probe radiance-sharing light probe generation on the scene \textit{Living Room}.}
    \label{fig:light_compare}
\end{figure}

\subsubsection{Visualization of Intermediate Results}

\textbf{Transfer coefficients.} In Figure \ref{fig:sh_compare}, we compare the ground-truth (GT) transfer coefficients with our predictions for three selected spherical harmonic channels: $(\ell=0,m=0),(\ell=1,m=0),(\ell=2,m=0)$. Our predicted coefficients closely match the ground truth, as shown by the small differences in most regions. The error map reveals that the main discrepancies appear on glossy surfaces, which exhibit high-frequency signals that are more challenging for the network to capture in full detail.

\textbf{Light Coeffieicnts.} In Figure \ref{fig:light_compare}, we compare the light coefficients interpolated using plain probe generation (where each probe's radiance map is independent) and our across-probe radiance-sharing light probe generation detailed in Section \ref{subsec:probe_interpolation}. For visualization, we selected three spherical harmonic channels: $(\ell=0,m=0),(\ell=1,m=0),(\ell=2,m=0)$. The results show that light coefficients interpolated using the proposed across-probe radiance-sharing strategy closely match those generated using plain probe generation for $\ell=1$, which accounts for a large portion of the illuminance. However, the differences become more pronounced for higher-frequency spherical harmonic channels, which indicates that the shared strategy loses accuracy in finer high-frequency spherical harmonics.

\begin{table}[t]
\centering
\caption{Quantitative comparison in terms of global illumination on the selected 6 scenes from the Bitterli dataset \cite{resources16}.}
\begin{tabular}{llccc}
\toprule
Scene & Method & RMSE & L1 & SSIM \\
\midrule
\multirowcell{2}{Bathroom}
& Ours & \textbf{0.108} & \textbf{0.087} & \textbf{0.741} \\
& DDGI & 0.282 & 0.244 & 0.611 \\
\multirowcell{2}{Dining Room}
& Ours & \textbf{0.033} & \textbf{0.014} & \textbf{0.901} \\
& DDGI & 0.061 & 0.037 & 0.504 \\
\multirowcell{2}{Living Room}
& Ours & \textbf{0.069} & \textbf{0.039} & \textbf{0.839} \\
& DDGI & 0.185 & 0.144 & 0.442 \\
\multirowcell{2}{Living Room 2}
& Ours & \textbf{0.075} & \textbf{0.053} & \textbf{0.815} \\
& DDGI & 0.269 & 0.212 & 0.655 \\
\multirowcell{2}{Living Room 3}
& Ours & \textbf{0.058} & \textbf{0.038} & \textbf{0.923} \\
& DDGI & 0.261 & 0.244 & 0.720 \\
\multirowcell{2}{Staicase}
& Ours & \textbf{0.039} & \textbf{0.029} & \textbf{0.878} \\
& DDGI & 0.185 & 0.130 & 0.165 \\
\bottomrule
\end{tabular}

\label{tab:gi_comparison}
\end{table}

\begin{table}[t]
\centering
\caption{Quantitative comparison in terms of global illumination+direct illumination on the selected 6 scenes from the Bitterli dataset \cite{resources16}.}
\begin{tabular}{llccc}
\toprule
Scene & Method & RMSE & L1 & SSIM \\
\midrule
\multirowcell{2}{Bathroom}
& Ours+DI & \textbf{0.096} & \textbf{0.067} & \textbf{0.836} \\
& DDGI+DI & 0.223 & 0.179 & 0.741 \\
\multirowcell{2}{Dining Room}
& Ours+DI & \textbf{0.023} & \textbf{0.010} & \textbf{0.960} \\
& DDGI+DI & 0.054 & 0.028 & 0.844 \\
\multirowcell{2}{Living Room}
& Ours+DI & \textbf{0.058} & \textbf{0.029} & \textbf{0.905} \\
& DDGI+DI & 0.132 & 0.094 & 0.756 \\
\multirowcell{2}{Living Room 2}
& Ours+DI & \textbf{0.059} & \textbf{0.037} & \textbf{0.897} \\
& DDGI+DI & 0.214 & 0.163 & 0.795 \\
\multirowcell{2}{Living Room 3}
& Ours+DI & \textbf{0.043} & \textbf{0.025} & \textbf{0.963} \\
& DDGI+DI & 0.119 & 0.100 & 0.900 \\
\multirowcell{2}{Staicase}
& Ours+DI & \textbf{0.034} & \textbf{0.025} & \textbf{0.902} \\
& DDGI+DI & 0.128 & 0.093 & 0.386 \\
\bottomrule
\end{tabular}

\label{tab:gt_comparison}
\end{table}

\subsubsection{Efficiency Evaluation}

We evaluate both temporal efficiency for training and inference stages and memory efficiency of our method.

\begin{figure}
    \centering
    \includegraphics[width=1\linewidth]{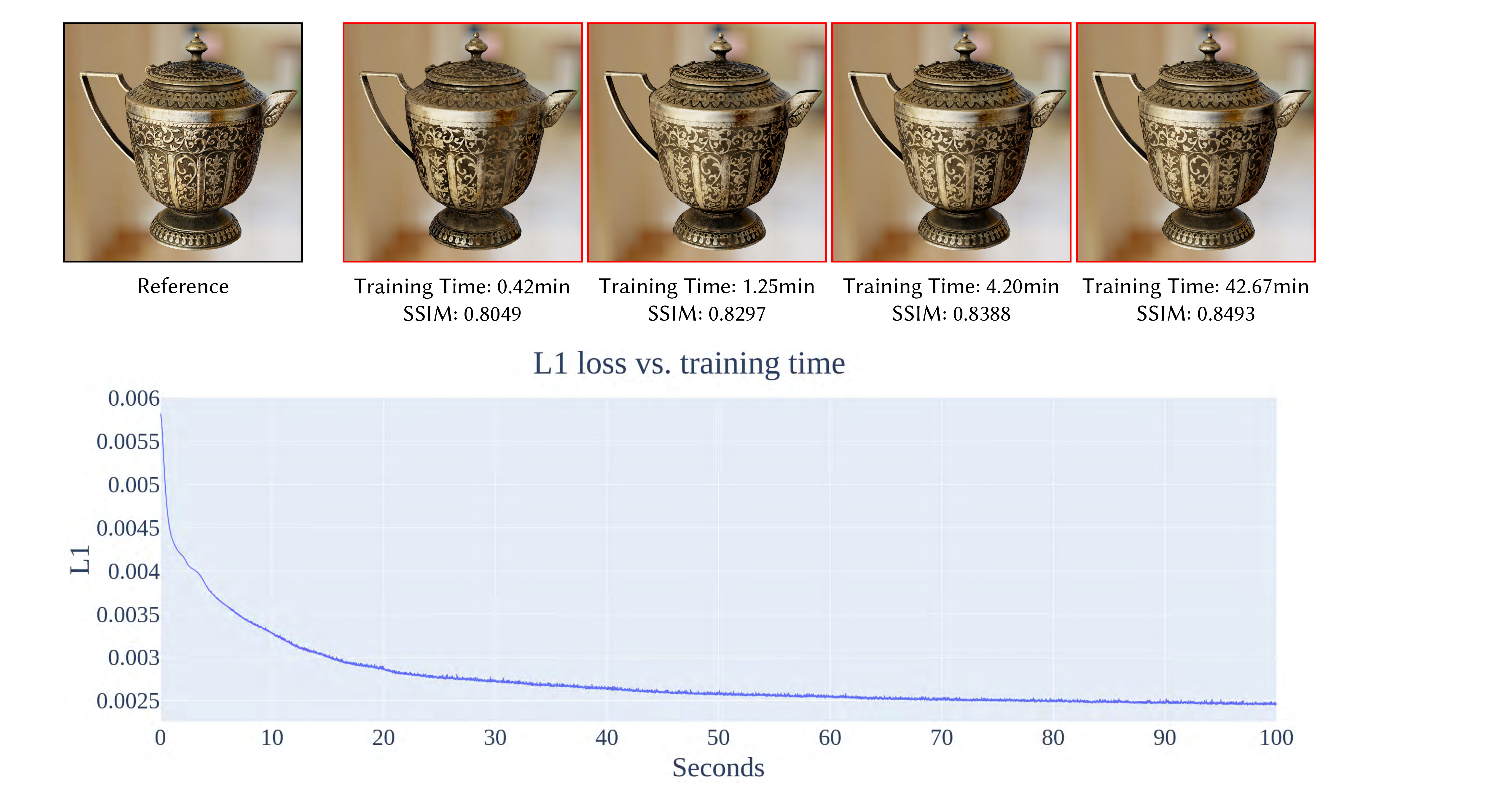}
    \caption{The temporal efficiency evaluation for training of our method.}
    \label{fig:trainingTime}
\end{figure}

The temporal efficiency evaluation for training of our method is shown in Figure \ref{fig:trainingTime}. The figure portrays the convergence curve of our method, showcasing its rapid convergence rate in achieving a relatively high SSIM metric.

The temporal efficiency evaluation for inference of our method in comparison with DDGI is presented in Table \ref{tab:efficiency}. This table showcases the performance metrics evaluated on the \textit{Living Room} scene, with the reported times corresponding to the rendering duration for a single frame. Overall, although DDGI is $4\times$ faster than our method, the time required to render a single frame with our approach is still less than 10ms, which satisfies the requirements of real-time rendering. DDGI achieves such a fast rendering speed due to its simplification of materials and low resolution of probe irradiance maps. Our method involves more stages, and each stage is quite independent, resulting in a longer rendering time for a single frame.

The memory usage of our method for the 6 evaluated scenes is presented in Table \ref{tab:memory}. We can observe that even for the scene with the largest number of vertices, \textit{Bathroom} with over 600k vertices, the memory footprint is less than 50Mb. This memory efficiency can be attributed to the fact that our default model appends only a 7-channel float vertex feature to each vertex. Consequently, our method can be highly memory-efficient, making it suitable for both off-the-shelf desktop devices and mobile devices with limited memory resources.

\begin{table}[t]
\centering
\caption{Efficiency evaluation of our method and DDGI on scene \textit{Living Room}.}
\begin{tabular}{llc}
\toprule
Method & Stage & Time(ms) \\
\midrule
\multirowcell{6}{Ours}
& Generate Network Input Buffer & 0.9 \\
& Network Inference & 1.5 \\
& Collect Shaded Points & 1.3 \\
& Probe Rasterization & 2.2 \\
& SH Multiplication(Render) & 0.8 \\
& \textbf{Total} & 6.7 \\
\midrule
\multirowcell{3}{DDGI}
& Probe Update & 1.2 \\
& Probe Interpolation(Render) & 0.5 \\
& \textbf{Total} & 1.7 \\
\bottomrule
\end{tabular}

\label{tab:efficiency}
\end{table}

\begin{table}[t]
\centering
\caption{Memory usage of our method for the 6 evaluated scenes from the Bitterli dataset \cite{resources16}.}
\resizebox{\linewidth}{!}{\begin{tabular}{lccc}
\toprule
Scene & Vertex Number & Network(Kb) & Vertex Features(Mb) \\
\midrule
Bathroom & 678985 & 799 & 42.33 \\
Dining Room & 141756 & 799 & 8.58 \\
Living Room & 118439 & 799 & 7.11 \\
Living Room-2 & 357509 & 799 & 21.77 \\
Living Room-3 & 451607 & 799 & 27.20 \\
Staircase & 179652 & 799 & 10.88 \\
\bottomrule
\end{tabular}
}
\label{tab:memory}
\end{table}

In summary, the proposed method \textit{TransGI} offers the following advantages:
\begin{itemize}
    \item Specular/glossy effects handling: Our method can handle ideal specular surfaces with an extra light bounce, and it effectively captures glossy effects within the representation capabilities of the spherical basis and the neural transfer model.
    \item Real-time rendering. Although not as efficient as DDGI \cite{majercik2019dynamic}, our method can still provide a real-time rendering experience with complex code divergence between CUDA and shader pipelines.
    \item Little storage overhead. 
\end{itemize}

\subsection{Ablation Study}
\label{subsec:ablation}

There are several hand-crafted parameters in our method, so to determine the optimal model parameters for my method, a series of ablation experiments are conducted, as shown in Table \ref{tab:ablation}, where metrics are evaluated on the scene \textit{Living Room} in terms of global illumination effect. The impact of each ablation choice is evaluated in terms of rendering quality (RMSE) rendering time changes (rendering time), along with the associated process. We explain the impact of our ablation choices below.

\textit{Resolution of irradiance map.} The resolution of the irradiance map for each probe significantly impacts the probe rasterization process. Higher resolutions result in more pixel-write operations and can improve rendering quality. However, there is a limit to the quality improvement achieved by increasing the resolution; quadrupling the resolution does not substantially enhance rendering quality. This is because the low-frequency global illumination is not sensitive to higher sampling rates. Conversely, downsampling the irradiance map resolution can fail to capture a wide range of shaded points, leading to degraded rendering quality. Therefore, an optimal resolution for the irradiance map must strike a balance. In this implementation, a resolution of 100 × 50 was chosen to achieve this balance.

\textit{Number of Probes.} It is evident that increasing the number of probes enhances rendering quality. This increase also results in a higher number of shaded points, thereby requiring more time for path tracing and probe rasterization. However, this does not necessarily translate to a linearly increased computation time if the additional workload remains within the parallel processing capabilities of the tested GPU. As shown in Table \ref{tab:ablation}, denser sampling of probes leads to increase in both rendering time and quality. Conversely, undersampling of probes results in a significant drop in rendering quality but not a significant drop in efficiency, which proves the default probe setup of $8\times8\times8$ is still within computational power of the tested GPU.

% \textit{Dimension of Vertex Features.} In theory, higher-dimensional vertex features have the capacity to encode more complex materials. However, my ablation experiments reveal that there is no significant difference in performance when the dimension of the vertex feature is set to either 7 or 23. (The actual number of dimensions is chosen to ensure that the concatenated feature dimension is a multiple of 16.) This observation suggests that a 7-dimensional vertex feature is enough for the evaluated scenes, whereas more complicated materials such as multi-layered materials will require higher-dimensional vertex features for representation. Furthermore, the number of input features only influences the computation of the first layer of the neural network, which explains why the rendering efficiency does not vary substantially between the two settings. The computational cost of the first layer is relatively small compared to the overall network, and the impact of changing the input feature dimension on the total rendering time is minimal.

\textit{Shaded rays per probe.} The number of shaded rays per probe for each frame is a scaling parameter that balances quality and efficiency. As shown in Table \ref{tab:ablation}, varying the number of shaded rays has minimal impact on rendering time and does not significantly affect rendering quality. This is primarily because the probes serve as a scene radiance cache that accumulates per-frame scene radiance, achieving a stable energy state within the initial few frames. Consequently, with sufficient frame accumulation, the probe energy stabilizes, making the number of shaded rays per probe per frame less critical.

\begin{table}[t]
\centering
\caption{Ablation study of the network hyperparameters' impact on the rendering quality and efficiency on the scene \textit{Living Room}.}
\label{tab:network_ablation}
\begin{tabular}{@{}ccc|cc@{}}
\toprule
\textbf{$d_{vertex}$} & \textbf{$d_{hidden}$} & \textbf{$l_{hidden}$} & \textbf{SSIM} & \textbf{Time(ms)} \\
\midrule
7 & 96  & 3 & 0.834 & 0.9 \\
7 & 128 & 3 & 0.839 & 1.5 \\
7 & 256 & 3 & 0.839 & 2.3 \\
23 & 128 & 3 & 0.842 & 1.5 \\
55 & 128 & 3 & 0.843 & 1.6 \\
7 & 128 & 4 & 0.846 & 2.0 \\
7 & 128 & 5 & 0.851 & 2.4 \\
\bottomrule
\end{tabular}
\end{table}

\textit{Hyperparameters of the neural transfer model.} There are many hyperparameters to be tuned to train the neural transfer model, including: the dimension of vertex feature $d_{vertex}$, the hidden dimension of the neural transfer model MLP $d_{hidden}$ and the number of hidden layers of the MLP $l_{hidden}$.
We provide a comprehensive ablation study of the above hyperparameters in Table \ref{tab:network_ablation}. From the table, we can see that by tuning the hyperparameters, we can achieve different trade-offs between rendering quality and efficiency. Increasing the hidden dimension ($d_{hidden}$) from 96 to 128 improves the SSIM from 0.834 to 0.839, but further increasing it to 256 yields no significant quality improvement while increasing the rendering time by 53\%. Similarly, increasing the vertex feature dimension ($d_{vertex}$) from 7 to 23 or 55 only marginally improves quality (0.839 to 0.843) with minimal impact on rendering time. The most effective quality improvement comes from increasing the number of hidden layers ($l_{hidden}$), with a 5-layer network achieving the highest SSIM of 0.851, though at a 60\% higher computational cost compared to the 3-layer network. Based on these results, we chose a balanced configuration with $d_{vertex}=7$, $d_{hidden}=128$, and $l_{hidden}=3$ for our experiments, which provides good quality while maintaining real-time performance.

\begin{table}[t]
\centering
\caption{Ablation experiment evaluations on quality and efficiency for global illumination of \textit{Living Room}.}
\resizebox{\linewidth}{!}{\begin{tabular}{lcc}
\toprule
Ablation Choice & \makecell{Time\\Changes(ms)} & SSIM \\
\midrule
Resolution of irradiance map =100$\times$200 & +6.0 & 0.842 \\
Resolution of irradiance map =50$\times$25 & -0.5 & 0.814 \\
Number of Probes =16$\times$16$\times$16 & +7.5 & 0.867 \\
Number of Probes =4$\times$4$\times$4 & -0.2 & 0.781 \\
Dimension of Vertex Features =23 & +0.2 & 0.841 \\
Shaded rays per probe =50 & -0.2 & 0.822 \\
Shaded rays per probe =200 & +1.5 & 0.853 \\
\midrule
Default Model & 6.7ms & 0.839 \\
\bottomrule
\end{tabular}
}
\label{tab:ablation}
\end{table}
\section{Limitations and Future Work}

% \textit{Fixed number of basis functions.} The proposed TransGI projects the transfer function into a frequency space represented by a chosen set of basis functions. The number of required basis functions varies greatly depending on the material properties, ranging from less than 10 for diffuse materials to hundreds for glossy materials and even infinitely many for specular materials, with more basis functions approximating the transfer function at higher computational costs. Our current implementation balances accuracy and efficiency by using a fixed number of 75 basis functions for all materials and employing an extra ray tracing step per pixel to approximate specular responses. Experiments in Figure\ref{fig:gi_comparison} show that \textit{TransGI} closely approximates ray tracing with high visual quality. However, its efficiency can be significantly improved by considering that the rendering environment is mainly composed of diffuse materials, which require far fewer basis functions. Future work will focus on dynamically determining the number of basis functions based on each material's transfer function to optimize the quality-efficiency trade-off.

\textit{Fixed Number of Basis Functions.} The proposed \textit{TransGI} projects the transfer function into a frequency space represented by a chosen set of basis functions. The requisite number of basis functions varies substantially, contingent upon the material properties. Diffuse materials necessitate fewer than 10 basis functions, while glossy materials demand hundreds, and specular materials require an infinite number. Our current implementation strikes a balance between accuracy and efficiency by employing a fixed number of 75 SH basis functions for all materials and utilizing an extra ray tracing step per pixel to approximate specular responses. Experiments in Figure \ref{fig:gi_comparison} demonstrate that \textit{TransGI} closely approximates ray tracing with high visual quality. Furthremore, the efficiency of our method can be significantly improved considering that the rendering environment predominantly comprises diffuse materials, which require far fewer basis functions. Future work will focus on dynamically determining the number of basis functions based on each material's transfer function to optimize the trade-off between quality and efficiency.

\begin{figure}
    \centering
    \includegraphics[width=0.9\linewidth]{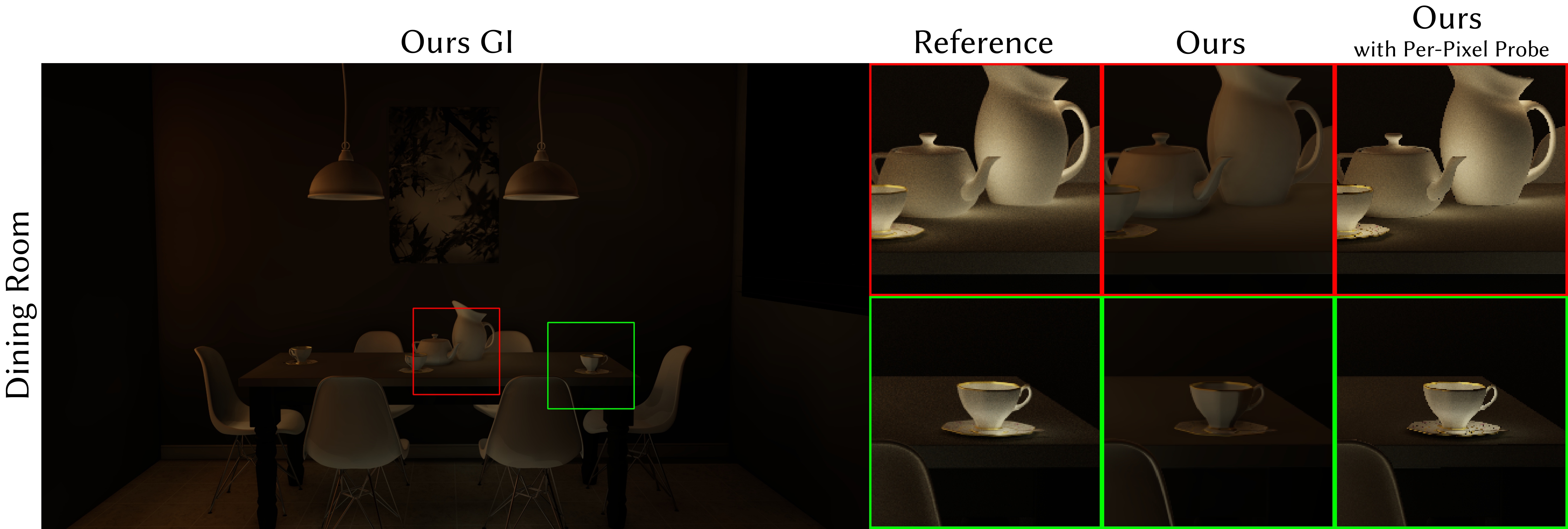}
    \caption{With sparse sh light probes, our method cannot produce accurate interreflection effects between two glossy objects.}
    \label{fig:interreflection}
\end{figure}

\textit{Object Interreflection.}
As illustrated in Figure \ref{fig:interreflection}, our method struggles to capture bright interreflection effects between two glossy objects. Traditional precomputed radiance transfer (PRT) techniques\cite{sloan2023precomputed} can model such effects by treating the entire scene as a whole and precomputing interreflected radiance transfer on each vertex. In contrast, our object-centric approach handles objects independently for dynamic rendering, neglecting inter-object interreflected radiance transfer. However, such interreflection effects could potentially be compensated by using denser light probes in our method. As shown in the rightmost column of Figure \ref{fig:interreflection}, we conducted an experiment with an extreme condition, placing light probes for each pixel, which reasonably restored the interreflection effects as observed in the reference path-tracing image. However, such a dense grid of light probes would lead to unacceptable computational costs, hindering real-time efficiency. Therefore, a future research direction would be to improve the efficiency of light probe generation or explore more efficient strategies to model interreflection effects between objects within our framework.

% \textit{Energy loss.} Compared with DDGI, our method can render quite illuminated global illumination effects, and the shadows are quite preserved with our decoded transfer coefficients. However, when compared with the path-traced reference image, our method still has a noticeable color difference. This is due to the loss of energy in both the appearance model and the light probe representation. The energy loss in the appearance model is mainly caused by the limited capacity and imperfect fitting of the neural network. The energy loss in the light probe representation is mainly caused by the limited number of light probes and the limited frequency range of the light coefficients. A future work could be to quantitatively analyze the energy loss in our method and propose to compensate for the residual to achieve better energy conservation.
\section{Conclusions}
\label{sec:conclusion}

In this paper, we have presented a novel global illumination technique that achieves real-time and high-fidelity rendering. Our key innovation is an object-centric neural transfer model that encodes the complex, multi-layered material properties of objects using a compact MLP and vertex-attached latent features. This lightweight, highly compressed representation facilitates real-time rendering of intricate objects with minimal memory overhead. Additionally, we have introduced radiance-sharing light probes to support local and dynamic illumination. By leveraging radiance information from neighboring probes, these probes increase the sample rate and quality of individual probes, yielding higher-fidelity results with reduced computation.

We have implemented our method in a real-time rendering engine, using compute shaders and CUDA-based neural networks, demonstrating its feasibility in practice. Our approach achieves real-time performance while delivering significantly improved rendering quality compared to baseline methods. In summary, our work presents a practical and efficient solution for real-time global illumination of neural rendering with complex, dynamic material properties, paving the way for advanced real-time rendering applications.

\bibliographystyle{IEEEtran}
\bibliography{template}

% \newpage

\section{Biography Section} 
\vspace{11pt}

\vspace{-33pt}
\begin{IEEEbiography}[{\includegraphics[width=1in,height=1.25in,clip,keepaspectratio]{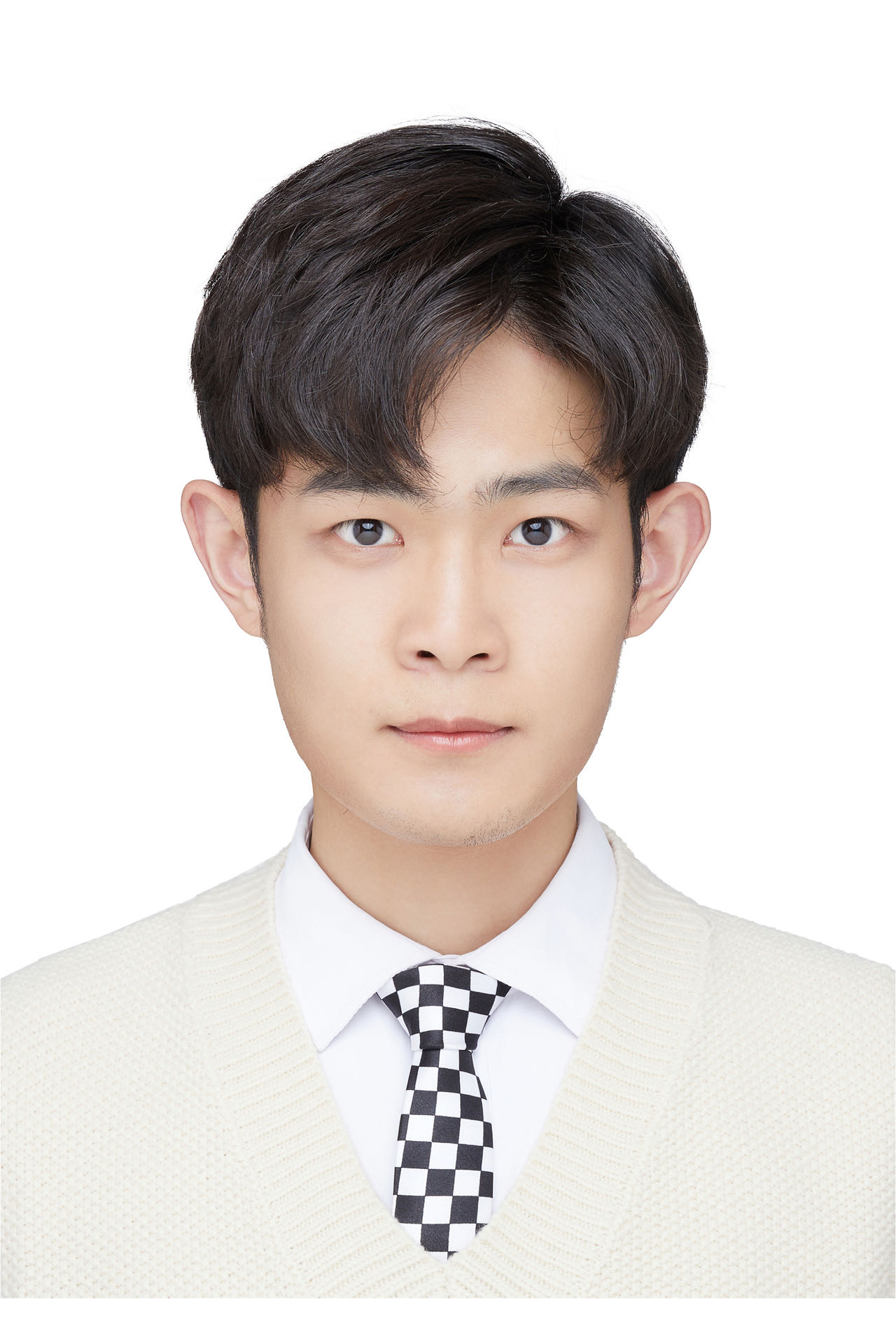}}]{Yijie Deng} is currently a PhD candidate student in New York University. He received Master's degree in Tsinghua-Berkeley Shenzhen Institute (TBSI), Tsinghua University in 2024. He received B.E. from Wuhan University in 2021. His research interest is 3D vision, Graphics and Robotics.
\end{IEEEbiography}

\vspace{11pt}

\vspace{-33pt}
\begin{IEEEbiography}[{\includegraphics[width=1in,height=1.25in,clip,keepaspectratio]{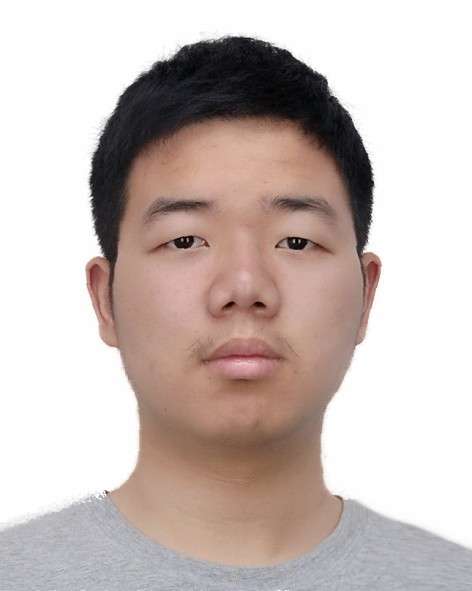}}]{Lei Han} is currently a researcher in Linx Lab of HiSilicon (subsidiary of Huawei Technologies). He studied Electrical Engineering at the Hong Kong University of Science and Technology and Tsinghua University. He received the B.S. degree in July 2013 and joined the Department of Electrical Computing Engineering at the Hong Kong University of Science and Technology in September 2016, where he is pursuing the PhD degree. His current research focuses on multi-view geometry and 3D computer vision.
\end{IEEEbiography}

\vspace{11pt}

\vspace{-33pt}
\begin{IEEEbiography}[{\includegraphics[width=1in,height=1.25in,clip,keepaspectratio]{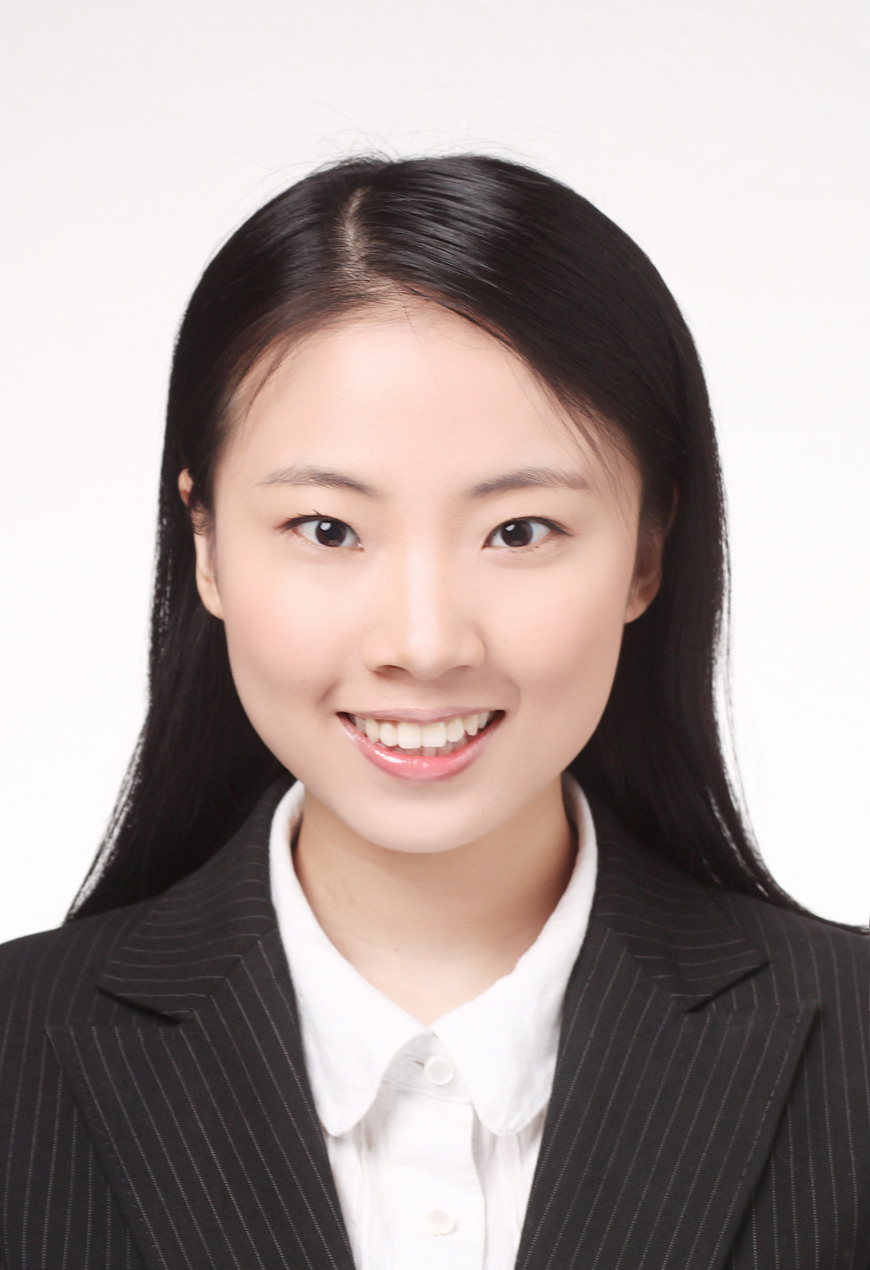}}]{Lu Fang} is currently an Associate Professor in Tsinghua University. She received Ph.D from the Hong Kong Univ. of Science and Technology in 2011, and B.E. from Univ. of Science and Technology of China in 2007. Her research interests include computational imaging and visual intelligence. Dr. Fang is currently IEEE Senior Member, Associate Editor of IEEE TIP and TMM.
\end{IEEEbiography}

\vfill

\end{document}